\documentclass[prl,twocolumn,superscriptaddress]{revtex4}
\usepackage{amssymb,amsmath,amsthm,graphicx}
\usepackage{latexsym}
\usepackage{bm}
\usepackage[]{hyperref}
\usepackage{color}

\allowdisplaybreaks

\pdfoutput=1

\begin{document}

\title{Localised $\bm{\mathrm{AdS}_3\times \mathrm{S}^3\times \mathbb{T}^4}$ Black Holes}

\author{ \'Oscar J. C. Dias}
\email{ojcd1r13@soton.ac.uk}
\affiliation{STAG Research Centre and Mathematical Sciences, University of Southampton, UK} 
\author{Jorge E. Santos}
\email{jss55@cam.ac.uk}
\affiliation{Department of Applied Mathematics and Theoretical Physics, University of Cambridge, Wilberforce Road, Cambridge CB3 0WA, UK} 

\begin{abstract}
We numerically construct asymptotically global $\mathrm{AdS}_3 \times \mathrm{S}^3 \times \mathbb{T}^4$ black holes in type IIB supergravity, with $\mathbb{R}_t \times SO(2) \times SO(3) \times U(1)^4$ symmetry, localised on the $\mathrm{S}^3$ and translationally invariant along the torus. These solutions, with horizons whose spatial cross sections have $\mathrm{S}^4 \times \mathbb{T}^4$ topology, dominate the microcanonical ensemble at low energies. At higher energies, a first-order phase transition occurs to $\mathrm{BTZ} \times \mathrm{S}^3 \times \mathbb{T}^4$ black holes — possessing $\mathbb{R}_t \times SO(2) \times SO(4) \times U(1)^4$ symmetry and $\mathrm{S}^1 \times \mathrm{S}^3 \times \mathbb{T}^4$ horizon topology. By the AdS/CFT correspondence, this transition reflects the spontaneous breaking of the $SO(4)$ R-symmetry of the D1-D5 CFT$_2$ to $SO(3)$. We also compute the expectation value of the scalar operator with lowest conformal dimension in the low-energy phase. Our $SO(3)$-localised black holes — together with the $U(1)^2$-localised solutions of \cite{Bena:2024gmp} — point to a rich landscape of novel black holes that may approach the $\mathrm{CFT}_2$ ‘sparseness bootstrap condition’, and shed light on how macroscopic entanglement in thermal phases encodes microscopic structure via internal directions.
\end{abstract}

\maketitle

{\bf~Introduction --} 
AdS/CFT correspondence is the prototypical example of a holographic duality \cite{Maldacena:1997re,Gubser:1998bc,Witten:1998qj} and remains a key tool for studying strongly coupled physics and quantum gravity. In particular,  D1-D5 CFT$_2$/AdS$_3$ duality provides a solvable testbed for exploring holography \cite{David:2002wn,Eberhardt:2019ywk}. On one side, the D1-D5 CFT$_2$, a two-dimensional $\mathcal{N}=(4,4)$ superconformal field theory, is among the simplest conformal field theories (CFTs). On the other, $\mathrm{AdS}_3 \times X$ gravity (with $X$ a compact manifold) is notably tractable — within pure gravity with global $\mathrm{AdS}_3$ asymptotics, the Ba\~nados-Teitelboim-Zanelli (BTZ) black hole is unique. When embedded in type IIB supergravity with $\mathrm{AdS}_3 \times \mathrm{S}^3 \times \mathbb{T}^4$ asymptotics, the $\mathrm{BTZ} \times \mathrm{S}^3 \times \mathbb{T}^4$ black hole (with sufficiently large temperature $T$) is dual to thermal states in the gauge theory on $\mathbb{R}_t \times \mathrm{S}^1$, with the same $T$ \footnote{We consider the low-energy limit with $N \to \infty$ and large 't Hooft coupling.}. Within this symmetry class, the type IIB equations of motion require the $\mathrm{S}^3$ radius $L$ to match the $\mathrm{AdS}_3$ length scale. These black holes preserve $\mathbb{R}_t \times SO(2) \times SO(4) \times U(1)^4$ symmetry and we shall henceforth refer to this phase simply as the BTZ, or uniform, phase. However, they only exist for energies $E/N^2 > 0$, while the dual CFT$_2$ has thermal states for $E>-\frac{c}{12}$, with $c = 6N^2$ the central charge. Hence, corresponding black holes should exist all the way down to $E = -\frac{c}{12}$.

This puzzle was partly resolved by \cite{Bena:2024gmp}, which analysed novel $\mathrm{AdS}_3 \times \mathrm{S}^3 \times \mathbb{T}^4$ black holes breaking $SO(4)$ symmetry to $U(1)^2$ — the so-called “$U(1)^2$ localised black holes” or “black poles” \footnote{These solutions were first constructed in \cite{Bah:2022pdn}, and were recently analysed in \cite{Sakamoto:2025jtn}.}. These exist for $-\frac{c}{12} < E \leq \frac{c}{96}$ and dominate the microcanonical ensemble for $-\frac{c}{12} < E < \frac{c}{24}(5\sqrt{5} - 11)$, partially overlapping with BTZ.

The existence of these $U(1)^2$ localised black holes hints at another class — those localised on $\mathrm{S}^3$, breaking $SO(4)$ down to $SO(3)$. Such `$SO(3)$ localised black holes' would be more symmetric than the $U(1)^2$ black poles, suggesting they should possess higher entropy at fixed energy. Unlike BTZ black holes and black poles — both with horizons whose spatial cross sections have $\mathrm{S}^1 \times \mathrm{S}^3 \times \mathbb{T}^4$ topology \footnote{Although the topology is the same in both cases, the geometry of the $S^3$ differs significantly. For the BTZ black hole, the $S^3$ is simply the standard round sphere, whereas for the black poles the $S^3$ incorporates the asymptotic $S^1$ contained within global AdS$_3$.} — $SO(3)$ localised black holes would have horizons with $\mathrm{S}^4 \times \mathbb{T}^4$ spatial cross sections.

One might expect $SO(3)$ localised black holes to exist by analogy with $SO(n)$-symmetric black holes in asymptotically $\mathrm{AdS}_k \times \mathrm{S}^n$ spacetimes \cite{Dias:2015pda,Dias:2016eto,Dias:2017uyv,Dias:2024vsc}, where the uniform phase becomes unstable to localisation via the Gregory-Laflamme mechanism \cite{Gregory:1993vy,Harmark:2002tr,Kol:2002xz,Kudoh:2004hs,Harmark:2007md,Headrick:2009pv,Dias:2015nua}. However, we find that the uniform $SO(4)$ phase — the $\mathrm{BTZ} \times \mathrm{S}^3 \times \mathbb{T}^4$ black hole—is not Gregory-Laflamme unstable. From this perspective, one might expect $SO(3)$ localised black holes not to exist.

Nevertheless, we were motivated to search for them, as the $\mathrm{S}^4 \times \mathbb{T}^4$ topology of the spatial cross sections of their horizon suggests they should appear point-like on $\mathrm{S}^3$ at low $E/c$, effectively behaving as six-dimensional spherical black holes with the torus as a spectator. Their entropy should scale as $S \sim (E + c/12)^{4/3}$, compared to $S \sim (E + c/12)^{3/2}$ for the $U(1)^2$ black poles \cite{Bena:2024gmp}, suggesting higher entropy at small energies. They should therefore dominate up to a critical energy, beyond which a first-order phase transition to either the BTZ black hole or a black pole may occur. By AdS/CFT, $\mathrm{S}^3$ symmetries correspond to R-symmetries in the D1-D5 CFT$_2$, so black holes breaking these symmetries are dual to thermal states with spontaneously broken R-symmetry (and nonzero scalar expectation values). This transition interpolates between symmetric thermal states at high energies and symmetry-breaking phases at low energies.

In this Letter, we numerically construct $SO(3)$ localised black holes and prove their existence. They always dominate over the $U(1)^2$ black poles of \cite{Bena:2024gmp} in the microcanonical ensemble. We also find they entropically dominate over the uniform phase (i.e. BTZ black hole) at low energies, determine the critical point of the first-order phase transition, and compute the expectation value of a scalar operator in the dual field theory. Via the duality, our results yield new quantitative predictions for the D1-D5 CFT$_2$.

{\bf~Problem Setup --} 
We focus on asymptotically $\mathrm{AdS}_3 \times \mathrm{S}^3 \times \mathbb{T}^4$ solutions in type IIB supergravity. The bosonic sector of type IIB supergravity can be succinctly described by the ten-dimensional metric $g_{10 {\rm D}}$, the complex axion–dilaton field $\tau =C_0+ i e^{-\phi}$, a self-dual five-form field strength $F_{(5)}$, and a complex three-form field strength $G_{(3)}$. We restrict attention to solutions where the axion–dilaton $\tau$ and the five-form field strength $F_{(5)}$ vanish. The ten-dimensional metric then takes the form
\begin{equation}
{\rm d}s^2_{10{\rm D}} = {\rm d}s^2_{6{\rm D}} + {\rm d}s^2_{\mathbb{T}^4},
\end{equation}
while the complex three-form field strength reduces to a Ramond–Ramond three-form that is self-dual with respect to the six-dimensional metric, {\emph i.e.} $G_{(3)} = F_{(3)} = {}_{6}\star F_{(3)}$. The vanishing dilaton, together with the self-duality of $F_{(3)}$, ensures that the numbers of D1 ($N_1$) and D5 ($N_5$) branes coincide, so that the central charge of the corresponding field theory is simply $c = 6 N_1 N_5 = 6 N^2$ \cite{Avery:2010qw}. Under these assumptions, the type IIB equations of motion reduce to
\begin{align} \label{eq:eom}
    &    E_{MN}\equiv R_{MN} - \frac{1}{4} F_{(3)\;MPQ} F_{(3)\;N}{}^{PQ}=0\;,\\
    &   {\rm d}F_{(3)}=0\;,\qquad 
        F_{(3)} ={}_{6}\star F_{(3)}\;,\nonumber
\end{align}
with uppercase Latin indices running over the six-dimensional spacetime. We seek static black hole solutions with topology $\mathrm S^4 \times \mathbb{T}^4$. Among these, the most symmetric configurations are expected to maximize the entropy. Such solutions exhibit $\mathbb R_t \times SO(2) \times SO(3) \times U(1)^4$ symmetry, preserving the full $SO(2)$ isometry of $\mathrm{AdS}_3$ together with the largest subgroup of $SO(4)$.

We employ the DeTurck method \cite{Headrick:2009pv,Dias:2015nua,Figueras:2011va}, which requires choosing a reference metric $\overline g$ that shares the symmetries and causal structure of the target solution.  We choose the reference metric
\begin{align}\label{eq:ansatzxy}
       \overline{\mathrm ds}^2&= \frac{L^2}{\left(1-y^2\right)^2} {\biggl \{} -H_1 {\mathrm d}t^2+\nonumber\\
       &\quad\qquad+H_2 \left[\frac{4\,\mathrm dy^2 }{2-y^2}+y^2 \left(2-y^2\right) \, {\mathrm d}\phi^2 \right]\biggr \}+  
        \\
& \quad  + L^2 H_2 {\biggl [} \frac{16\,\mathrm d x^2}{2-x^2} 
+4 x^2 \left(2-x^2\right) \left(1-x^2\right)^2\, {\mathrm d}\Omega_2^2  \biggr ], \nonumber
\end{align}
where $H_1$, $H_2$ are judiciously chosen functions. Specifically, when $H_1 = H_2 = 1$, the metric reduces to global $\mathrm{AdS}_3 \times \mathrm{S}^3$. Consequently, we require that $H_1$ and $H_2$ approach 1 as $y \to 1$ to recover this asymptotic geometry. Additionally, $H_1$ and $H_2$ must be chosen so that the reference metric describes a regular $\mathrm{S}^4$ black hole. In fact, for small black holes, the geometry near the horizon is expected to approximate the six-dimensional asymptotically flat Schwarzschild solution ($\mathrm{Schw}_6$).

To ensure that $H_1$ and $H_2$ satisfy these requirements, we apply the following coordinate transformations. First, we define $y = \sqrt{1 - \mathrm{sech}(Y)}$ and $x = \sqrt{1 - \sin(X/2)}$, which map the $\mathrm{d}x^2$ and $\mathrm{d}y^2$ components of the reference metric to $L^2 H_2 (\mathrm{d}X^2 + \mathrm{d}Y^2)$, a form that is conformal to Cartesian coordinates. Finally, setting $X = \rho \, \xi \, \sqrt{2 - \xi^2}$ and $Y = \rho \, (1 - \xi^2)$ defines a transformation from Cartesian to polar coordinates using an alternative polar coordinate set $\{\rho, \xi\}$. After the mapping, the DeTurck reference metric \eqref{eq:ansatzxy} becomes
\begin{align}\label{eq:ansatzrhoxi}
       \overline{\mathrm ds}^2&= -M\left(\frac{\rho^3-\rho_0^3}{\rho^3+\rho_0^3}\right)^2  {\mathrm d}t^2+L^2 H_2  {\biggl \{} {\mathrm d}\rho^2+\rho^2 {\biggl [} \frac{4{\mathrm d}\xi^2}{2-\xi ^2}+\nonumber\\   
&\qquad+G_1 \xi^2(2-\xi^2) {\mathrm d}\phi^2+G_2 \left(1-\xi ^2\right)^2  \, {\mathrm d}\Omega_2^2
{\biggr ]}  \biggr \},
\end{align}
where the map  $\{x,y\} =\{x(\rho,\xi),y(\rho,\xi)\}$  uniquely determines $G_1$ and $G_2$, and relates $M$ directly with $H_1$. In the $\mathrm{d}t^2$ component, the factor $(\rho^3 - \rho_0^3)^2 / (\rho^3 + \rho_0^3)^2$ is included to anticipate the placement of a black hole horizon in the reference metric. We set $H_2 = (1 + \rho_0^3 / \rho^3)^{4/3}$, which satisfies the requirement that $H_2 \to 1$ as $y \to 1$ ($\rho \to \infty$). For small $\rho$ and $\rho_0$, we have $G_1 \approx G_2 \approx 1$, so if $M \approx 1$, the reference metric approximates $\mathrm{Schw}_6$ in isotropic coordinates. This behavior guides our choice of $H_1$, and consequently of $M$ and $G_3$. We choose $H_1$ so that $H_1 = 1$ at $y \to 1$ ($\rho \to \infty$) and $M$ is positive definite with $M = 1$ at $\rho = \rho_0$. This condition also fixes the temperature and ensures the regularity of the horizon. Explicit expressions for these functions are provided in the Supplemental Material. 

Given a reference metric $\bar{g}$, the DeTurck method transforms the Einstein equation \eqref{eq:eom} into
\begin{equation}
E_{MN}-\nabla_{(M}\xi_{N)}=0\;,\qquad \xi^M \equiv g^{PQ}[\Gamma^M_{PQ}-\overline{\Gamma}^M_{PQ}]\;,
\end{equation}
where $\Gamma^M_{PQ}$ and $\overline{\Gamma}^M_{PQ}$ define the Levi-Civita connections for $g$ and $\bar g$, respectively. Unlike \eqref{eq:eom}, this formulation yields PDEs that are elliptic in nature. After solving these equations, we must check that $\xi^M = 0$ to ensure that \eqref{eq:eom} is satisfied. The local uniqueness of elliptic equations guarantees that solutions with $\xi^M = 0$ are distinct from those with $\xi^M \neq 0$. Moreover, the condition $\xi^M = 0$ completely fixes the coordinate freedom in the metric.

We adopt a general ansatz consistent with the symmetries, with the metric depending on six arbitrary functions to be determined numerically. We now turn our attention to $F_{(3)}$ Locally, the three-form field strength $F_{(3)}$ can be expressed in terms of a two-form potential $C_{(2)}$ as $F_{(3)} = {\rm d}C_{(2)}$, since ${\rm d}F_{(3)} = 0$. The gauge potential $C_{(2)}$ takes the form
\begin{equation}
C_{(2)}= L^2 F \,{\mathrm d}t  \wedge {\mathrm d}\phi+L^2 W \, {\rm Vol}_{\Omega_2}\;,
\end{equation}
where $F$ and $W$ are unknown functions, and ${\rm Vol}_{\Omega_2}$ is the volume form on $S^2$. Applying the self-duality condition together with the Bianchi identity for $F_{(3)}$ leads to a single second-order PDE for $F$, which allows $W$ to be fully determined in terms of $F$ and its derivatives, and thus eliminated from all the equations of motion.

Our integration domain contains five boundaries: the horizon $\rho=\rho_0$, asymptotic infinity $y=1$ ($\rho\to\infty)$, the $\mathrm{S}^1$ axis $y=0$ ($\xi=0$), the $\mathrm{S}^3$ `North' pole $x=1$ ($\xi=1$), and the `South' pole $x=0$. At infinity, as a boundary condition, we impose global $\mathrm{AdS}_3\times \mathrm{S}^3$ asymptotics. Regularity determines the remaining boundary conditions (details in Supplemental Material). 

To handle the five boundaries numerically, we divide the integration domain into six warped rectangular `patches' (see Supplemental Material).  The four patches $I-IV$ far from the horizon use $\{x,y\}$ coordinates, while the remaining two patches $V-VI$ use $\{\rho,\xi\}$ coordinates.  We require that the metric $g$, the form $C_{(2)}$, and their first derivatives match across patch boundaries.  

We therefore have a boundary value problem for seven functions of two variables. $L$ drops out of the equations of motion, so the only parameter is $\rho_0$ which fixes the temperature \footnote{Our reference metric limits $\rho_0<\pi$, but our solutions are well within this bound.}. We solve the system using a Newton-Raphson algorithm, starting from the reference metric and $F = 1$ at $\rho_0 = 0.1$ as the initial seed. Pseudospectral collocation is employed, with transfinite interpolation of Chebyshev grids in each patch, and the resulting linear systems are solved via LU decomposition. All solutions satisfy $\xi^2 < 10^{-10}$. Further details, including explicit expressions for our ansatz and numerical convergence tests, are provided in the Supplemental Material (see also \cite{Dias:2015nua}).

\begin{figure}[ht]
\centering
\includegraphics[width=.35\textwidth]{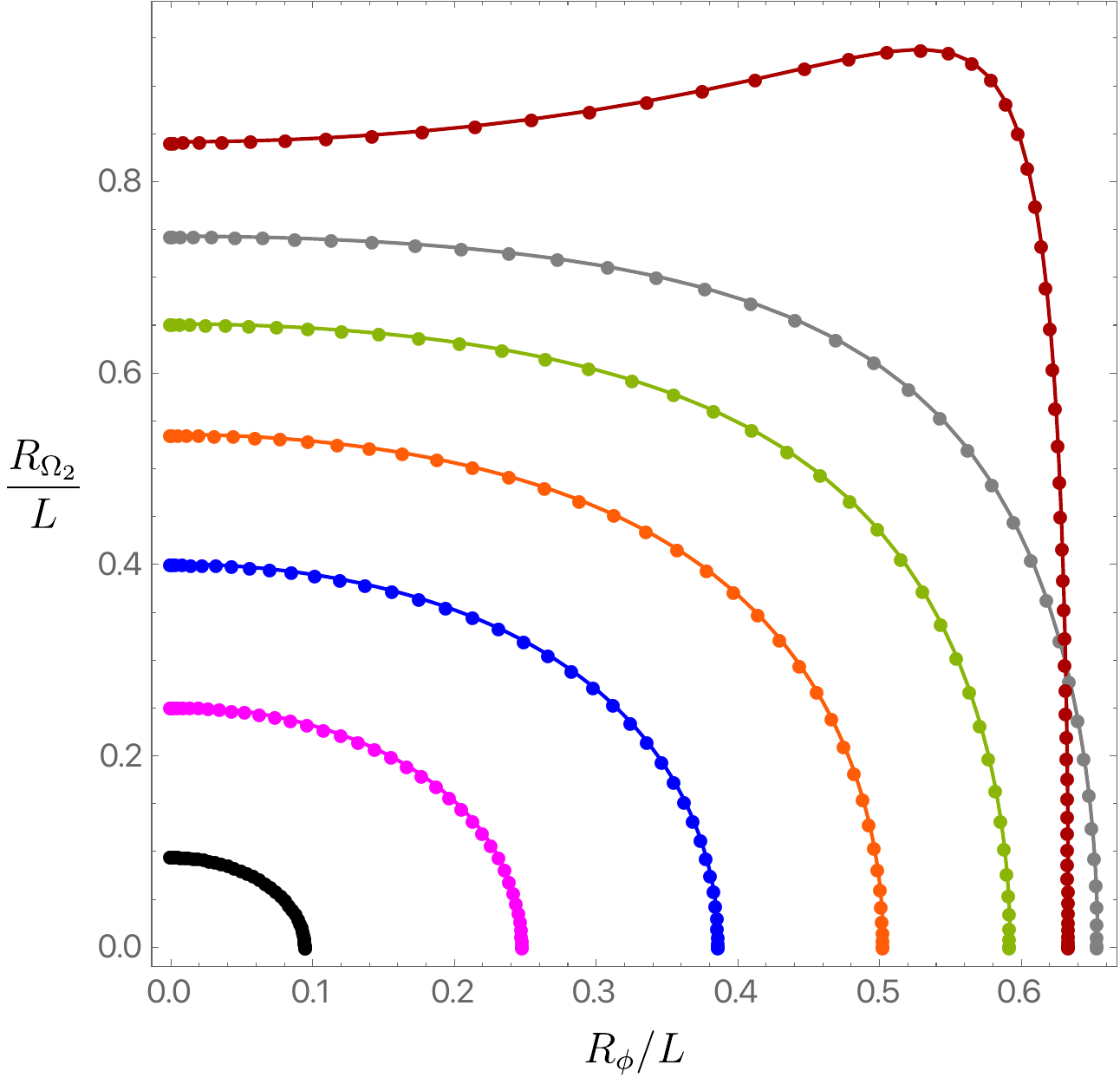}
\caption{Radii of the $\mathrm{S}^1$ and $\mathrm{S}^2$ (in units of $L$) along the horizon. From bottom-left to top-right the curves have $T L\approx \{2.507,0.940,0.578,0.418,0.327,0.269,0.123\}$.}\label{Fig:deformation}
\end{figure}

{\bf~Results --} 
Fig.~\ref{Fig:deformation} presents a parametric plot of the radii $R_{\phi}$ and $R_{\Omega_2}$ of the geometrically preserved $\mathrm{S}^1$ and $\mathrm{S}^2$ along the horizon. For small $\rho_0$ (high temperatures), the curve is well approximated by $R_{\phi}^2 + R_{\Omega_2}^2 \approx 2^{4/3} \, \rho_0^2 \, L^2$, indicating that the horizon is nearly spherical. As $\rho_0$ increases (lower temperatures), the horizon becomes significantly more deformed.

We now compute the thermodynamic quantities. The temperature $T$ is determined by $\rho_0$, while the entropy $S$ is obtained by integrating the horizon area. The energy $E$ is calculated using either the Kaluza-Klein holographic renormalization formalism \cite{Kanitscheider:2006zf,Kanitscheider:2007wq} (see also \cite{Skenderis:2006uy,Dias:2015pda,Kim:1985ez,Gunaydin:1984fk,Lee:1998bxa,Lee:1999pj,Arutyunov:1999en,Skenderis:2006di,Skenderis:2007yb}) or the covariant Noether charge (covariant phase) method \cite{wald1993black, iyer1994some, iyer1995comparison, wald2000general, papadimitriou2005thermodynamics, Dias:2019wof} (see also the Supplemental Material). In the AdS/CFT dictionary, the six- and three-dimensional Newton constants are related to the central charge of the dual CFT$_2$ via $G_6 = \frac{\pi^2}{2} \frac{L^4}{N^2}$ and $G_3 = \frac{G_6}{2\pi^2 L^3}$. Numerically, these thermodynamic quantities satisfy the first law, $\mathrm{d}E = T,\mathrm{d}S$, to better than $0.1\%$ accuracy.  

\begin{figure}[ht]
\centering
\includegraphics[width=.4\textwidth]{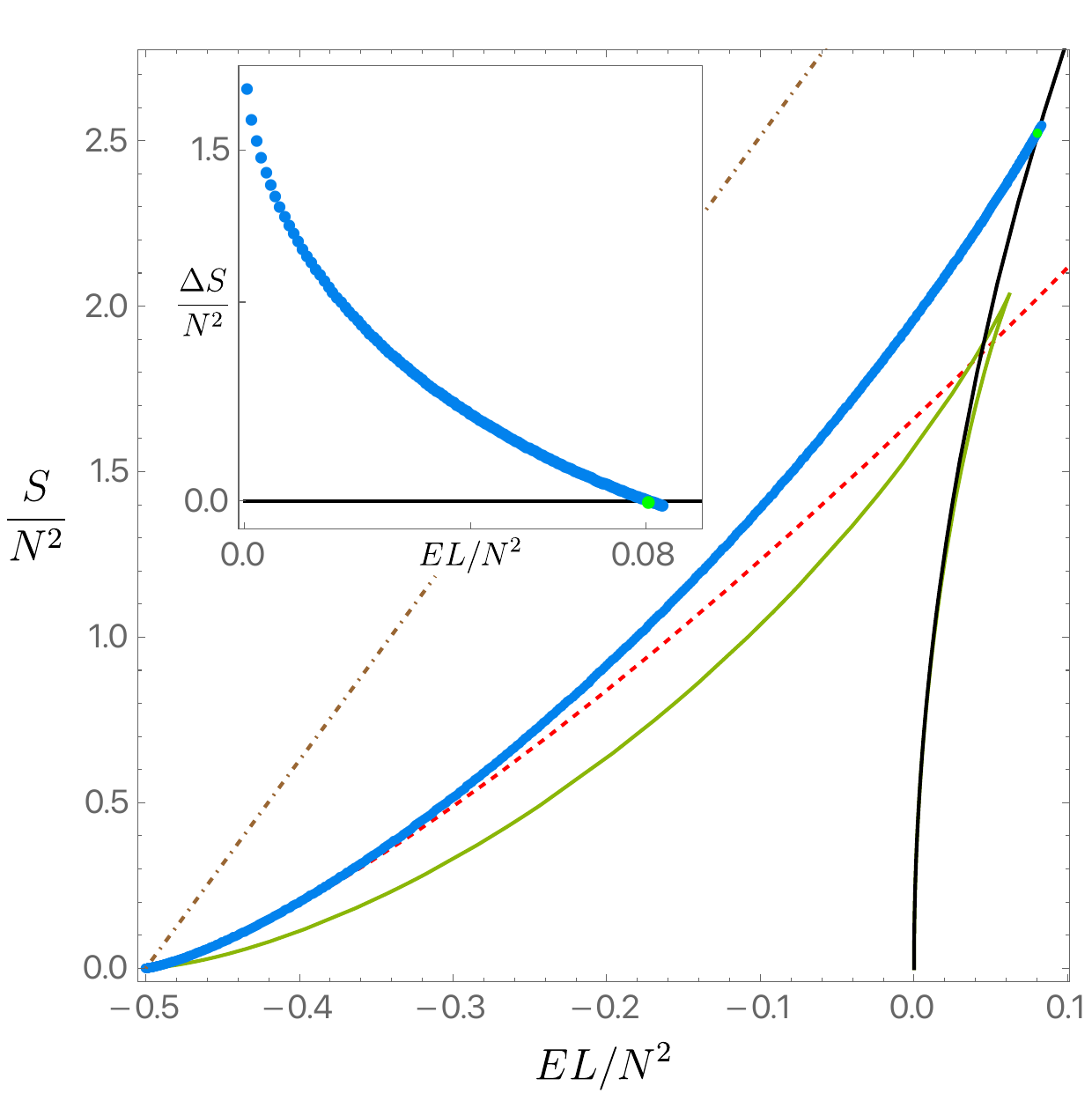}
\caption{Microcanonical phase diagram: entropy vs energy. The black line is the $\mathrm{BTZ}\times \mathrm{S}^3\times \mathbb{T}^4$ phase, the green curve is the $U(1)^2$ localised phase \cite{Bena:2024gmp} and the blue dots are the $SO(3)$ localised phase. The red dashed line is the lowest-order $\mathrm{Schw}_{6}\times \mathbb{T}^4$ approximation and the dot-dashed brown line is the `sparseness' bound.}\label{Fig:microcanonical}
\end{figure}

In the microcanonical ensemble, the energy is fixed, and the dominant solution is the one that maximizes the entropy. Fig.~\ref{Fig:microcanonical} shows $S/N^2$ versus $E L/N^2$ for various competing solutions. At low energies, the entropy of the $SO(3)$-localised black hole (blue disks) is well approximated by that of $\mathrm{Schw}_6 \times \mathbb{T}^4$ (dashed red curve; hereafter, we restore the spectator torus). When both solutions coexist, starting at $E L/N^2 = -1/2$, the $SO(3)$-localised black holes always have higher entropy than the $U(1)^2$-localised solutions of \cite{Bena:2024gmp} (green curve). The inset plot shows the entropy of the $SO(3)$-localised solutions relative to that of $\mathrm{BTZ} \times \mathrm{S}^3 \times \mathbb{T}^4$ (continuous black curve) in the coexistence region. For energies $- \frac{1}{2} \frac{N^2}{L} \leq E < E_c$, the $SO(3)$-localised black holes remain the phase with the highest entropy, exceeding even that of $\mathrm{BTZ} \times \mathrm{S}^3 \times \mathbb{T}^4$. At the critical energy $E_c L/N^2 \approx 0.0804\pm0.0002$ (with $S_c/N^2 \approx 2.52216\pm0.0002$, indicated by the green point), a first-order phase transition occurs: above $E_c$, the $\mathrm{BTZ} \times \mathrm{S}^3 \times \mathbb{T}^4$ solution dominates the microcanonical ensemble, having the highest entropy. 

Returning to the discussion in the Introduction, we note that, unlike the uniform solutions in other $\mathrm{AdS}_k \times \mathrm{S}^n$ systems (e.g., $\mathrm{AdS}_5 \times \mathrm{S}^5$ \cite{Hubeny:2002xn,Dias:2015pda,Buchel:2015gxa,Dias:2016eto,Bosch:2017ccw,Dias:2017uyv,Dias:2024vsc}), we find that $\mathrm{BTZ} \times \mathrm{S}^3 \times \mathbb{T}^4$ is {\it not} unstable to a Gregory-Laflamme–type instability at any energy. Remarkably, this does not rule out the existence of $SO(3)$-localised solutions. However, it does exclude the existence of ‘lumpy’ black holes — solutions with horizon topology $\mathrm{S}^1 \times \mathrm{S}^3 \times \mathbb{T}^4$ that are deformed along the polar direction of $\mathrm{S}^3$ — which are perturbatively close to BTZ. Such lumpy black holes do exist in the $\mathrm{AdS}_5 \times \mathrm{S}^5$ system \cite{Dias:2015pda,Dias:2016eto,Dias:2017uyv,Dias:2024vsc}.

To see this explicitly, we consider linear perturbations of $\mathrm{BTZ} \times \mathrm{S}^3 \times \mathbb{T}^4$ at temperature $T$. We perform a Fourier decomposition $e^{-i \omega t}$ along the time direction $t$ and a harmonic decomposition on the $S^2$, expanding all perturbations in terms of the standard spherical harmonics $Y_{\ell m}$ with $\ell \in \mathbb{Z}^+$ and $|m| \le \ell$. Solving the resulting eigenvalue problem, we find (details to be presented elsewhere) that the frequencies are quantised as:
 \begin{equation} \label{NoGL:freq}
 \omega= -2\pi T\,(2+2p+\ell)\,i\,,
 \end{equation}
where $p\in \mathbb{Z}^+$ is a radial overtone. Since these frequencies always have a negative imaginary part, there is no Gregory-Laflamme instability.

As noted in the Introduction, localised black holes are dual to thermal states in which the R-symmetry of the D1-D5 CFT$_2$ is spontaneously broken, in our case down to $SO(3)$. This breaking triggers the condensation of an infinite tower of scalar operators with increasing conformal dimension $\Delta$, the lowest of which has dimension $2$. The expectation value $\langle \mathcal{O}_2\rangle$ of this operator in the broken phase can be computed using the formalism of Kaluza–Klein holography \cite{Kanitscheider:2006zf,Kanitscheider:2007wq} (see Supplemental Material). Fig.~\ref{fig:expectation} shows $\langle \mathcal{O}_2\rangle$ across a range of energies. Because the symmetry-breaking transition is first-order, $\langle \mathcal{O}_2\rangle$ takes a nonzero value at the phase transition.

\begin{figure}[ht]
\centering
\includegraphics[width=.4\textwidth]{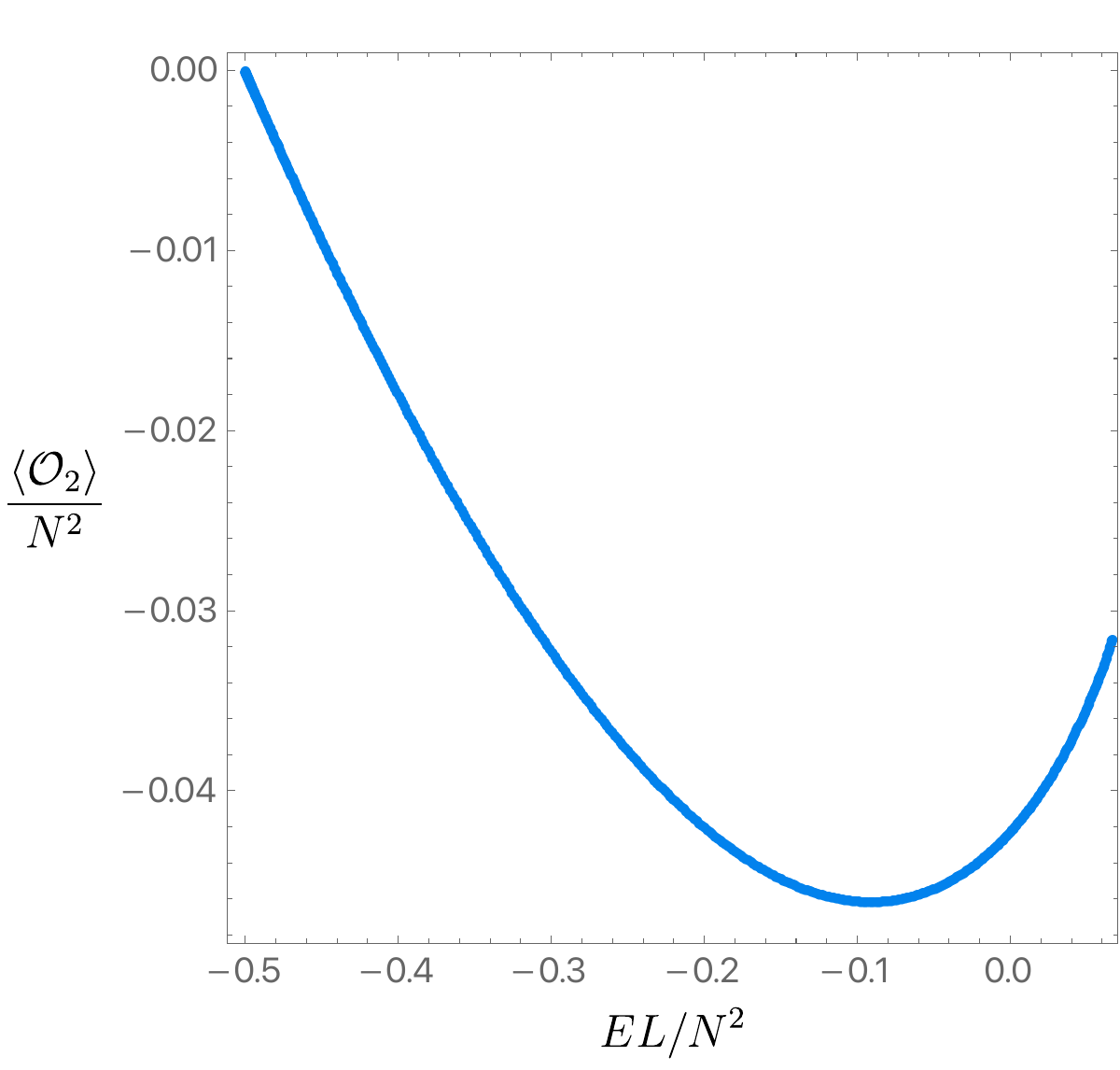}
\caption{Dimension 2 scalar condensate vs energy. 
}\label{fig:expectation}
\end{figure}   
 
In the canonical ensemble, the temperature is fixed, and the solution with the lowest free energy $F=E-TS$ dominates. In this ensemble, a first-order Hawking–Page phase transition occurs between thermal $\mathrm{AdS}_3\times \mathrm{S}^3\times \mathbb{T}^4$ at low temperatures and $\mathrm{BTZ}\times \mathrm{S}^3\times \mathbb{T}^4$ black holes at high temperatures \cite{Hawking:1982dh,Witten:1998zw}. Localised black holes, whether $SO(3)$ or $U(1)^2$, never dominate in this ensemble (see the Supplemental Material for a phase diagram).
 
{\bf~Discussion --} We have numerically constructed asymptotically global $\mathrm{AdS}_3\times \mathrm{S}^3\times \mathbb{T}^4$ localised black holes in type IIB supergravity that preserve the largest subgroup $SO(3)$ of $SO(4)$. These black holes, with horizon topology $\mathrm{S}^4 \times \mathbb{T}^4$, are more entropic than any other known solutions at low energies, with a first-order transition to $\mathrm{BTZ}\times \mathrm{S}^3\times \mathbb{T}^4$ black holes near $E L/N^2\approx 0.0804$. Via AdS/CFT, they are dual to thermal states of the strongly coupled D1-D5 CFT$_2$ with spontaneously broken R-symmetry, where the lowest-dimension scalar operator has $\Delta=2$ and preserves $SO(3)$ (see Fig.~\ref{fig:expectation}).

Lattice studies of holographic field theories have focused on the canonical ensemble \cite{Catterall:2007fp,Anagnostopoulos:2007fw,Catterall:2008yz,Hanada:2008ez,Catterall:2009xn,Hanada:2011fq,Hanada882}, but recent work on first-order transitions in other ensembles \cite{PhysRevLett.68.9, PhysRevLett.71.211, PhysRevE.64.056101} suggests that the D1-D5 CFT$_2$ at large $N$ and strong coupling should reproduce Figs.~\ref{Fig:microcanonical} and \ref{fig:expectation}, providing then a precision holography test.

$SO(3)$ Localised solution on the $S^3$ with two or more black holes like ours should exist (very much like in the original black string setup \cite{Horowitz:2002dc,Harmark:2003eg,Harmark:2003yz,Dias:2007hg}). They should exist even with different local masses but should be dynamically unstable to merging into our single $SO(3)$ Localised black hole and thus should have less entropy. Such infinite set of multi-black hole solutions should fill the area below the blue curve and the black curve (and slightly extend to the right of the latter) in the phase diagram of Fig.~\ref{Fig:microcanonical}. 

The phase diagram that completes the $SO(3)$-Localised curve in Fig.~\ref{Fig:microcanonical} can be conjectured: analogous to the green $U(1)^2$ solutions, $SO(3)$ black holes likely extend to a maximum energy, forming a lower branch down to $\{ E,S \}=\{0,0\}$ where they meet extremal $\mathrm{BTZ}\times \mathrm{S}^3\times \mathbb{T}^4$ and extremal $U(1)^2$ holes (see Supplemental Material).

Modular bootstrap constraints on large-$c$ CFTs \cite{Hartman:2014oaa,Bena:2024gmp} impose a sparseness condition, $S(E)\lesssim 2\pi \left( E+ \frac{c}{12} \right)$ for $-\frac{c}{12}\leq E \leq \frac{c}{12}$, with $c=6N^2$ for D1-D5 CFT$_2$. The $SO(3)$ branch lies well below this bound (see dot-dashed brown curve in Fig.~\ref{Fig:microcanonical}). Solutions approaching the bound might require localisation on $\mathbb{T}^4$, generalising our translationally invariant solutions \cite{Dias:2017coo}. This would solve the  puzzle raised by the CFT sparsness bound. Solutions breaking more $\mathrm{S}^3$ symmetries likely have lower entropy.

Computing entanglement entropy $S_{EE}$ and other observables for $SO(3)$ black holes would test conjectures from \cite{Bena:2024gmp}. There, $U(1)^2$ and $\mathrm{BTZ}\times \mathrm{S}^3\times \mathbb{T}^4$ phases have similar $S_{EE}$ at small energies due to an “entanglement shadow” where Ryu–Takayanagi surface \cite{Ryu:2006bv,Hubeny:2007xt} cannot probe infrared differences \cite{Balasubramanian:2014sra,Caminiti:2024ctd,Arora:2024edk}. $SO(3)$ solutions, with distinct horizon topology, may deviate from this universal behaviour and could provide non-trivial tests of the eigenstate thermalisation hypothesis \cite{Srednicki:1994mfb,PhysRevA.43.2046}, including examples of quantum scars \cite{Bernien:2017ubn,Turner:2018kjz}. They also provide a platform to apply novel methods for computing entanglement in solutions with non-trivial internal dependence \cite{Bena:2024gmp}.
Finally, $\mathrm{BTZ}\times \mathrm{S}^3\times \mathbb{T}^4$ is the near-horizon limit of the D1-D5-P black hole. AdS/CFT reproduces its microscopic entropy \cite{Strominger:1996sh}, and asymptotically $\mathrm{AdS}_3\times \mathrm{S}^3\times \mathbb{T}^4$ solutions are central to the fuzzball program \cite{Mathur:2005zp,Skenderis:2008qn,Mathur:2024ify}. Understanding the thermodynamics and microscopic description of $SO(3)$ localised black holes could provide valuable insights into this framework.

\vskip .2cm
\centerline{\bf Acknowledgements}
\vskip .2cm
We are grateful to Iosif~Bena and Pierre~Heidmann for their valuable comments on an earlier draft and for stimulating discussions that enriched our results. O.~J.~C.~D. acknowledges financial support from the STFC “Particle Physics Grants Panel (PPGP) 2020” Grant No. ST/X000583/1.  J.~E.~S. has been partially supported by STFC consolidated grant ST/X000664/1. J.~E.~S. would also like to thank the Simons Center for Geometry and Physics for its hospitality, where this work was completed. The authors acknowledge the use of the IRIDIS High Performance Computing Facility, and associated support services at the University of Southampton, in the completion of
this work. 
\onecolumngrid  \vspace{1cm} 
\begin{center}  
{\Large\bf Appendix} 
\end{center} 
\appendix 


\section{Action and equations of motion}
We are interested on solutions of IIB supergravity with $\mathrm{AdS}_3\times \mathrm{S}^3\times \mathbb{T}^4 $ asymptotics.  We restrict our attention to cases where the only mode of the compact torus $\mathbb{T}^4$ that can be excited is the breathing mode (which only deforms the overall volume of $\mathbb{T}^4$). Instead of working with 10-dimensional IIB, we can then do a dimensional reduction along $\mathbb{T}^4$. We obtain 6-dimensional supergravity which is a consistent truncation of IIB supergravity in the sense that all solutions of the former can be uplifted to 10 dimensions and are necessarily a solution of IIB (see e.g. \cite{Mathur:2005zp,Kanitscheider:2006zf,Kanitscheider:2007wq,Skenderis:2008qn,Mathur:2024ify} and references therein).
The action of type IIB supergravity (in the string frame) for the sector of solutions of interest here is
\begin{equation}  \label{S10}
S_{10}=\frac{1}{16\pi G_{10}}\int \mathrm{d}^{10} x \sqrt{-g_{10}}\left[ e^{-2\Phi}\Big( R_{10} +4(\nabla\Phi)^2 \Big)-\frac{1}{12} F_{(3)}^2 \right],
\end{equation}
where $G_{10}$ is 10-dimensional Newton's constant,  $g$, $\Phi$ and $F_{(3)}$ are the graviton, dilaton and RR 3-form $F_{(3)}=\mathrm{d}C_{(2)}$ and we have set to zero all other IIB fields that are not excited in the solutions we look for ($R_{10}$ is the Ricci scalar). The 3-form gauge field sources both electric D1 brane and magnetic D5 brane charges (recall that, by electromagnetic duality, the Hodge star of a 7-form is a 3-form, $\star F_{(7)}=F_{(3)}$). The metric (in the string frame) used in the reduction takes the following form
\begin{equation}
{\rm d}s^2_{10{\rm D}} = {\rm d}s^2_{6{\rm D}} + e^{\Phi}{\rm d}s^2_{\mathbb{T}^4},
\end{equation}
where the radius of the $\mathbb{T}^4$ is related to the dilaton $\Phi$.

The resulting reduced action of 6-dimensional supergravity (in the Einstein frame) is then
 \begin{equation}  \label{S6}
S_{6}=\frac{1}{16\pi G_{6}}\int \mathrm{d}^{6} x \sqrt{-g}\left[ R -(\nabla\Phi)^2 -\frac{1}{12} e^{2\Phi}F_{(3)}^2 \right],
\end{equation}
where $G_{6}=G_{10}/(2^4\pi^4 V_4)$  and $R_{6}$ are the 6-dimensional Newton's constant and Ricci scalar ($(2\pi)^4V_4$ is the $\mathbb{T}^4$ volume). The 6-dimensional metric in the Einstein frame is precisely the 6-dimensional part of the 10-dimensional metric (in the string frame) because the 6-dimensional dilaton, $\Phi_6 = \Phi - \frac{1}{4}\ln \mathrm{det} \,g_{\mathbb{T}^4}$, is a constant. Further note that if the 10-dimensional solution were to depend on the internal torus directions, under dimensional reduction one would end up with a (possibly infinite) tower of 6-dimensional extra fields (as stated above we do not consider this case here).
The 6-dimensional equations of motion that follow from \eqref{S6} are (in the trace-reversed form):
\begin{subequations}\label{eq:eom6D}
\begin{align} 
        R_{MN} - \frac{1}{4} e^{2\Phi} \left( F_{(3)\;MPQ} F_{(3)\;N}{}^{PQ} -\frac{1}{6}g_{MN} F_{(3)}^2\right) - \nabla_M\Phi \nabla_N\Phi &=0\,\\
        {\rm d}\star (e^{2\Phi}F_{(3)})&=0\,,\\
        \Box \Phi &=\frac{1}{12} e^{2\Phi} F_{(3)}^2\,.
\end{align}
\end{subequations}
When the number of D1 and D5 branes is the same, $N_1=N_5 \equiv N$, the dilaton vanishes, $\Phi=0$, and the 3-form is self-dual, $\star F_{(3)}=F_{(3)}$ ($\Rightarrow F_{(3)}^2=0$). Then, \eqref{eq:eom6D} reduce to the equations of motion \eqref{eq:eom} in the main text.     

\section{Ansatz for $SO(3)$ localised black holes}

We want to find asymptotically $\mathrm{AdS}_3\times \mathrm{S}^3\times \mathbb{T}^4 $ solutions of type IIB supergravity that are static and have horizon topology $\mathrm S^4\times \mathbb{T}^4$. These $SO(3)$ localised black holes preserve the largest subgroup $SO(3)$ of the $SO(4)$ symmetry group of $\mathrm{S}^3$. Furthermore, we are interested on solutions that are translational invariant along the torus so we solve directly the Einstein-DeTurck equations of motion of 6-dimensional supergravity and keep the torus fixed and undeformed.

Within the Einstein-DeTurck formution of the equations of motion \cite{Headrick:2009pv,Dias:2015nua,Figueras:2011va}, a DeTurck reference metric tailored to find such solutions must therefore contain an axis, a topologically $S^4$ horizon, and asymptote to ${\mathbb R}^{(1,2)}\times \mathrm{S}^3$  (within 6-dimensional supergravity). That is, it must contain the same causal structure and symmetries as the localised black hole solution we search for. So, asymptotically it must approach $\mathrm{AdS}_3\times \mathrm{S}^3$. On the other hand, for very small energies, the solution near the horizon is expected to resemble the 6-dimensional Schwarzschild-Tangherlini with a round $\mathrm{S}^4$. But as the energy increases one expects that the $S^4$ gets increasingly deformed (as best shown in Fig.~\ref{Fig:deformation} of the main text).
 To accommodate the five physical boundaries of the system, it is convenient to work with two different patches, each one with four boundaries (including non-physical patching boundaries) and their own coordinate system. Four of these patches ($I-IV$) and associated coordinate chart are adapted to the asymptotic region, and the other two ($V-VI$) to the near horizon region.
 
Next, we give  in full our ansatz for the $SO(3)$ localised black holes and associated DeTurck reference metric in both coordinate charts. 
 The ansatz in the $\{x,y\}$ far-region coordinate system is given by
\begin{subequations}\label{eqAPP:ansatzxy}
\begin{align}
       \mathrm ds^2&= \frac{L^2}{\left(1-y^2\right)^2} {\biggl \{} - H_1 f_1 {\mathrm d}t^2+H_2 \left[\frac{4 f_2}{2-y^2}{\mathrm d}y^2+  y^2 \left(2-y^2\right) f_3 \, {\mathrm d}\phi^2 \right]\biggr \}+  \nonumber \\
& \quad\qquad  + L^2 H_2 {\biggl [} \frac{16}{2-x^2} f_4 \left({\mathrm d}x+ f_6 {\mathrm d}y\right)^2 +4 x^2 \left(2-x^2\right) \left(1-x^2\right)^2 f_5 \, {\mathrm d}\Omega_2^2  \biggr ]\;,\\
          C_{(2)}&=L^2 \frac{y^2 \left(2-y^2\right)}{\left(1-y^2\right)^2} H_1 f_7 \,{\mathrm d}t  \wedge {\mathrm d}\phi+L^2 W \,\mathrm {\rm Vol}_{\Omega_2} \;,
\end{align}
\end{subequations}
where $f_i$ are unknown functions to be determined numerically, the remaining quantities are known and given below, and ${\rm Vol}_{\Omega_2}$ is the volume form on $S^2$. The asymptotic boundary is at $y=1$, the $\mathrm{S}^1_{\phi}$ axis is at $y=0$ and the North and South poles of $\mathrm{S}^3$ are at $x=0$ and $x=1$, respectively.
The DeTurck reference metric can be recovered by setting $f_i=1$ for $i\neq 6$ and $f_6=0$, matching the one presented in the main text. It matches global $\mathrm{AdS}_3\times \mathrm{S}^3$ asymptotically if we choose $H_1$ and $H_2$ to be such that $H_1\to 1$ and $H_2\to 1$ at $y=1$ to recover global $\mathrm{AdS}_3\times \mathrm{S}^3$ asymptotically.  $H_1$ and $H_2$ must also be chosen so that the reference metric describes a regular $\mathrm S^4$ black hole (more below). 

After performing the polar $\{x,y\}$ to Cartesian $\{X,Y\}$ coordinate transformation followed by the Cartesian $\{X,Y\}$ to polar $\{\rho,\xi\}$ coordinate change described in the main text, one gets the coordinate transformation between the far-region coordinates $\{x,y\}$ and the near-region coordinates $\{\rho,\xi\}$. This is: 
\begin{subequations} \label{APPchartmap} 
\begin{align}
	y&=\sqrt{1-{\rm sech} \left(\rho \,\xi \sqrt{2-\xi^2} \right)}\,,\qquad\qquad x=\sqrt{1-\sin \left[\frac{1}{2} \,\rho \left(1-\xi ^2\right)\right]}\;,\\
	\rho&=\sqrt{{\rm arcsech}\left(1-y^2\right)^2+4\,{\rm arcsin}\left(1-x^2\right)^2}\,,\qquad\qquad \xi=\sqrt{1-\frac{2 {\rm arcsin}\left(1-x^2\right)}{\sqrt{{\rm arcsech}\left(1-y^2\right)^2+4\,{\rm arcsin}\left(1-x^2\right)^2}}}\;.
\end{align}
\end{subequations}


In the $\{\rho,\xi\}$ near-region coordinate system, our ansatz reads
\begin{subequations}\label{eqAPP:ansatzrhoxi}
\begin{align}
       {\mathrm d}s^2&= -L^2 M {f}_1 \frac{\left(\rho^3-\rho_0^3\right)^2}{\left(\rho^3+\rho_0^3\right)^2}  {\mathrm d}t^2+L^2 H_2  {\biggl \{}  \tilde{f}_2  {\mathrm d}\rho^2+\rho^2 {\biggl [} \frac{4\tilde{f}_4 (  {\mathrm d}\xi+ \tilde{f}_6 {\mathrm d}\rho)^2}{2-\xi ^2}+G_1 \xi^2(2-\xi^2){f}_3 {\mathrm d}\phi^2+G_2 \left(1-\xi ^2\right)^2 {f}_5 \, {\mathrm d}\Omega_2^2
{\biggr ]}  \biggr \}, \\
          C_{(2)}&= L^2 \xi^2 \left(2-\xi ^2\right) \rho^2 \frac{\left(\rho^3-\rho_0^3\right)^2}{\left(\rho^3+\rho_0^3\right)^2} M G_3 {f}_7 \,{\mathrm d}t  \wedge {\mathrm d}\phi+L^2 W \,{\rm Vol}_{\Omega_2}\;,
\end{align}
\end{subequations}
where $\tilde f_2$, $\tilde f_4$, and $\tilde f_6$ are new unknown functions.  The remaining functions transform between the far and near  coordinate systems as scalars via \eqref{APPchartmap}. On the DeTurck reference metric, these functions are $\tilde f_2=1$, $\tilde f_4=1$, and $\tilde f_6=0$, matching the one presented in the main text. If, for small $\rho$ and $\rho_0$, we have $G_1\approx G_2\approx 1$, and $M\approx1$, the reference metric does approximate $\mathrm{Schw}_{6}$ in isotropic coordinates for the appropriate choice of $H_2$.  This further guides our choice for $H_1$ (and consequently for $M$ and $G_3$) and $H_2$ (more below). In \eqref{eqAPP:ansatzrhoxi}, the horizon is at $\rho=\rho_0$, and the Equator and North pole of $\mathrm{S}^4$ are at $\xi=0$ and $\xi=1$, respectively. 

In \eqref{eqAPP:ansatzxy} and \eqref{eqAPP:ansatzrhoxi}, the known functions (judiciously chosen to obey the conditions that we have been listing) are given by
\begin{subequations}
\begin{align}
	G_1&={\rm sinc}\left(i \,\rho\,\xi  \sqrt{2-\xi ^2} \right)^2, \quad G_2={\rm sinc}{\biggl [}\rho \left(1-\xi ^2\right) {\biggr ]}^2,  \quad G_3={\rm sinc}\left(i \,\rho \,\xi  \sqrt{2-\xi ^2} \right)^2 {\rm sech}\left(\rho \,\xi  \sqrt{2-\xi ^2} \right)^2\\
        H_1&=\frac{E_{-}^2}{E_{+}^2}\left[\left(1-y^2\right)^2 + y^2\left(2-y^2\right)\frac{E_{-}^2}{E_{+}^2}\right]\;,\\
        H_2&=\left(1+\frac{\rho_0^3}{\rho^3}\right)^{4/3}=\left\{1+\frac{\rho_0^3}{\left[{\rm arcsech}\left(1-y^2\right)^2+4\,{\rm arcsin}\left(1-x^2\right)^2\right]^{3/2}}\right\}^{4/3}\;,\\
        M&=
1+\frac{\left(\rho_0^3-\rho^3\right)^2}{\left(\rho^3+\rho_0^3\right)^2}\sinh^2\left(\rho\,\xi \sqrt{2-\xi ^2} \right)\;,
\end{align}
\end{subequations}

where
\begin{equation}
E_{\pm}=\rho_0^3 \pm \left[{\rm arcsech}\left( 1-y^2\right)^2+4\,{\rm arcsin}\left(1-x^2\right)^2 \right]^{3/2}\;.
\end{equation}
and the function ${\rm sinc} z=\frac{\sin z}{z}$ for $z\neq 0$ and ${\rm sinc} z=1$ for $z=0$. 

The Einstein-de Turck equations motions $-$ \eqref{eq:eom} in the main text $-$ for \eqref{eqAPP:ansatzxy} and \eqref{eqAPP:ansatzrhoxi} must be solved together with the boundary conditions that we now discuss. At the asymptotic boundary $y=1$ the solution must approach the reference metric and thus we impose Dirichlet conditions $f_{1,2,3,4,5,7}=1$ together with
\begin{equation}
\left.\bigg( f_6+\frac{2-x^2}{32}\frac{\partial^2f_4}{\partial x\partial y}\bigg) \right|_{y=1}=0\,.
\end{equation}

Regularity at the $\mathrm{S}^1_{\phi}$ axis ($y=0$) require that $f_2=f_3, f_6=0$ and that the other functions obey Neumann boundary conditions. At the North ($x=0$) and South ($x=1$) poles of $\mathrm{S}^3$ regularity of the solution requires that $f_4=f_5, f_6=0$ and Neumann boundary conditions for the other functions at $x=0$ or mixed boundary conditions at $x=1$ (that follow straightforwardly from the equations of motion at $x=1$). 
At the horizon $\rho=\rho_0$, regularity of the solution  requires that $\tilde{f}_1=\tilde{f}_2, \tilde{f}_6=0$ and that the other $\tilde{f}_i$ obey certain Robin boundary conditions (whose expressions  are long and unilluminating). 
At the Equator ($\xi=0$) regularity of the solution requires that we impose $\tilde f_3=\tilde f_4, \tilde f_6=0$ and Neumann boundary conditions for the other functions $\tilde f_i$. Finally, at the North pole ($\xi=1$) of $\mathrm{S}^4$ regularity demands that one must have $\tilde f_4=\tilde f_5, \tilde f_6=0$ and Neumann boundary conditions for the other functions $\tilde f_i$. Regularity of solutions at asymptotic boundaries, horizons and axes of symmetry are discussed in detail in section V of \cite{Dias:2015nua} and in  section 3 of \cite{Dias:2010maa}.
There is a detail that should be highlighted. In the ansatz \eqref{eqAPP:ansatzxy} and \eqref{eqAPP:ansatzrhoxi}, there is a cross term which, in the $\{x,y\}$ coordinates, can be written schematically as $\sim(1-y^2)^p f_6\mathrm dx \mathrm dy$, for some power $p$.  On the reference metric one has $f_6=0$, so the reference metric is unaffected by $p$.  However, our boundary condition at infinity $(y=1)$ \emph{is} affected by the power $p$, since it determines the particular fall-off of the cross term. Therefore, the choice of $p$ does hold physical significance and our choice of $p=0$ is such that the various operators in the dual field theory are unsourced.  

\begin{figure}[ht]
\centering
\includegraphics[width=.45\textwidth]{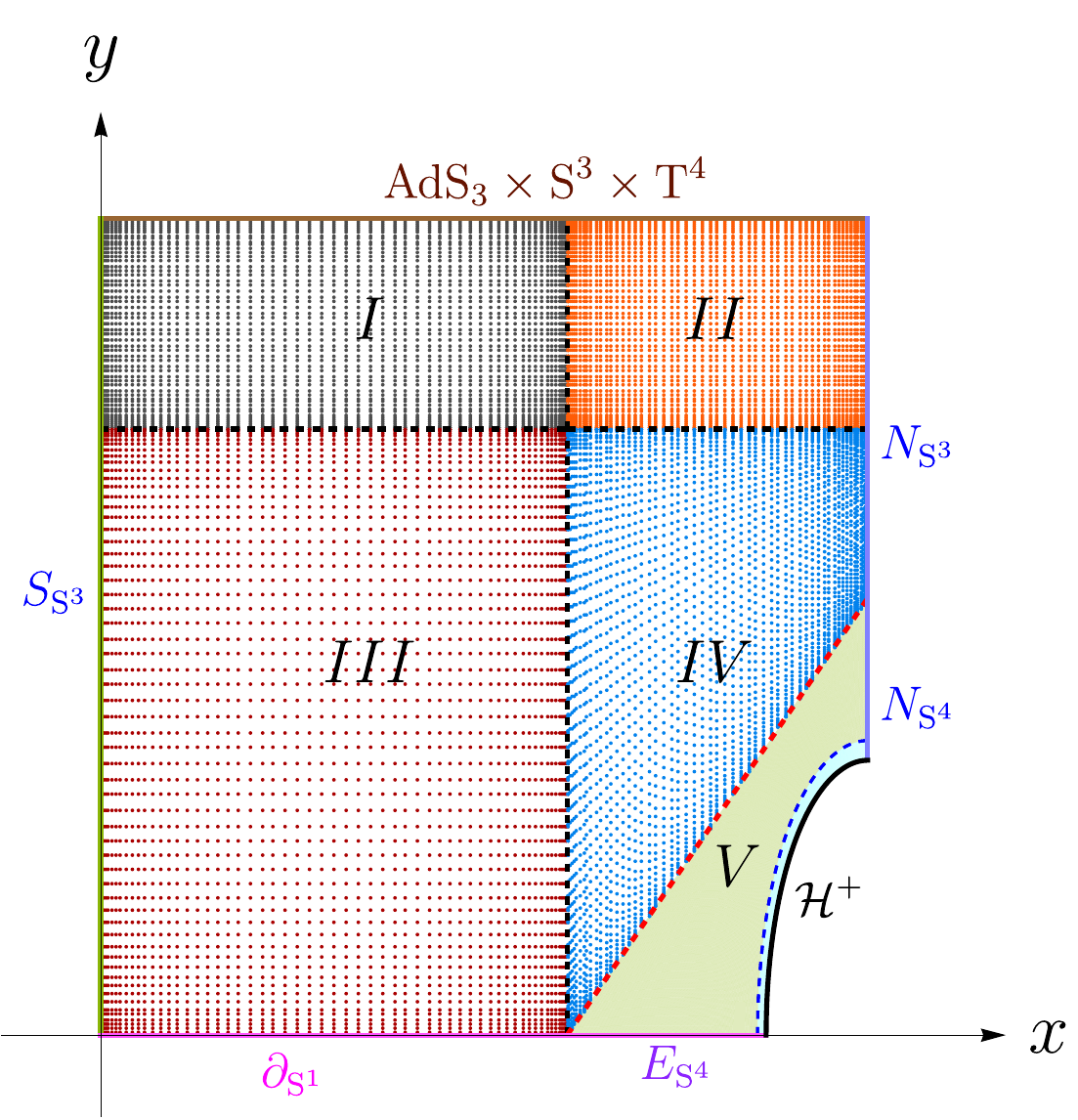}
\hspace{1cm}
\includegraphics[width=.47\textwidth]{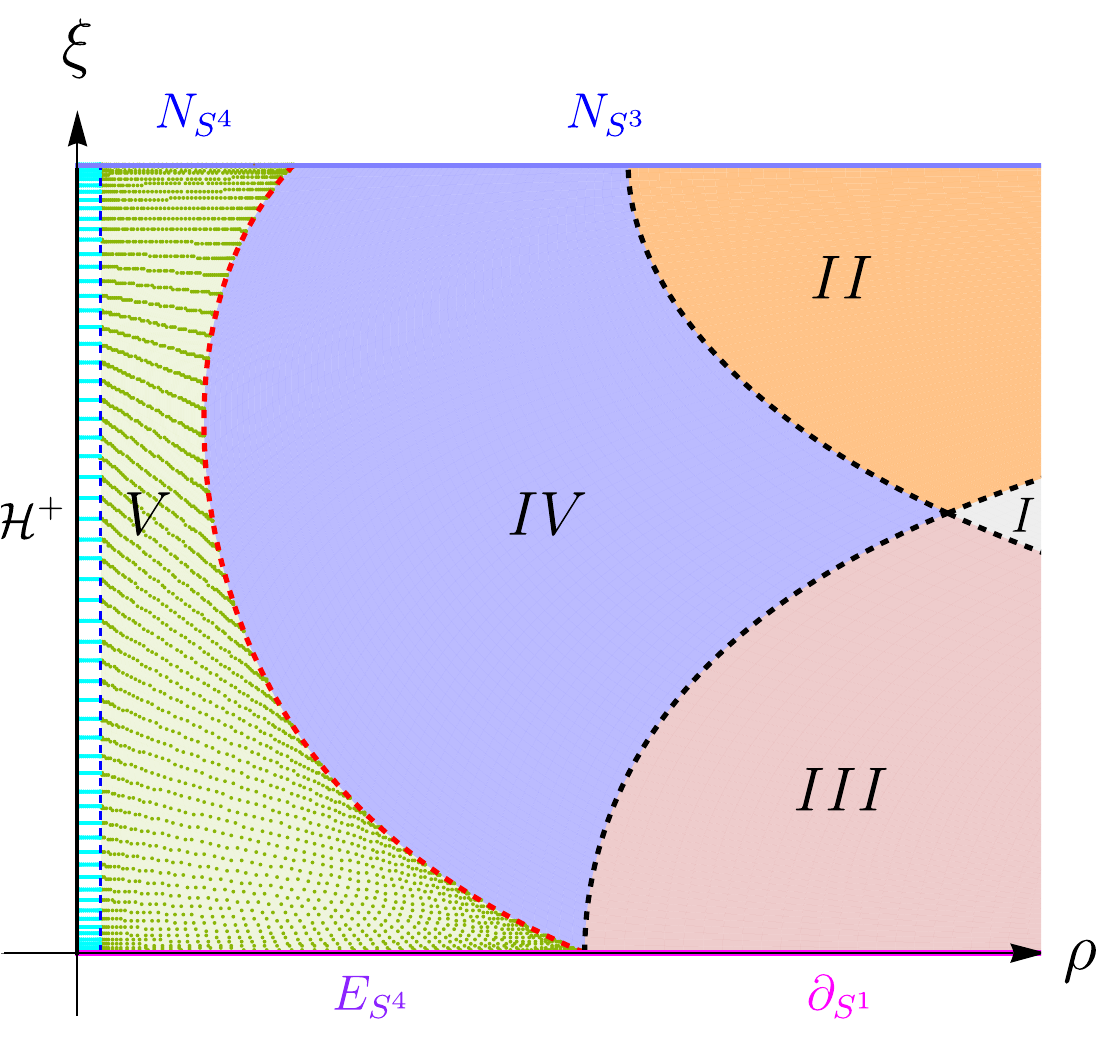}
\caption{Integration domain with 6 patches $I-VI$ (small patch $VI$ near the horizon $\mathcal{H}^+$ has no label in figure).  Chebyshev-Gauss-Lobatto grids with $ \widetilde{N}\times \widetilde{N}$ points are placed using transfinite interpolation. {\bf Left panel:} Patches $I-IV$ (in the far region) use  $\{x,y\}$ coordinates and patches $V-VI$ (near the horizon) are mapped from $\{\rho,\xi\}$ coordinates using \eqref{APPchartmap}.  {\bf Right panel:} Patches $V-VI$ (near the horizon) use  $\{\rho,\xi\}$ coordinates. In the far-region the four physical boundaries are the asymptotic boundary $\mathrm{AdS}_3\times \mathrm{S}^3\times \mathbb{T}^4 $, the North ($N_{\mathrm{S}^3}$) and South ($S_{\mathrm{S}^3}$) poles of the $\mathrm{S}^3$ and the rotation axis of $\mathrm{S}^1$ of $\mathrm{AdS}_3$. In the near-horizon the three physical boundaries are the horizon $\mathcal{H}^+$ and the North ($N_{\mathrm{S}^4}$) and south ($S_{\mathrm{S}^4}$) poles of the $\mathrm{S}^4$. In plots figures one has $\rho_0=0.5, \rho_1=0.54, k_0\simeq  0.608,  f_0\simeq 0.742, p_0\simeq 0.535$.}\label{Fig:patches}
\end{figure}

We solve the boundary value problem numerically using a Newton-Raphson algorithm.  Typically, to discretise the numerical grid,  we divide the integration domain into six non-overlapping warped rectangular regions or `patches'  as shown in Fig.~\ref{Fig:patches}: in the far region patches $I-IV$ we use  $\{x,y\}$  coordinates while in the near horizon patches $V-VI$ we use  $\{\rho,\xi\}$ coordinates. In each patch we place Chebyshev-Gauss-Lobatto $N \times N$ grids using transfinite interpolation (these methods are reviewed in \cite{Dias:2015nua}; the results we present have up to $N=70$). The patching boundary between patches $I-III$ and $II-IV$ is given by $y=f_0$ and between $I-II$ and $III-IV$ it is given by $x=k_0$. The patching boundary between patches IV and V (magenta curve) is given by $y=\frac{p_0}{1-k_0}(x-k_0)$. Finally, the patch boundary between patches $V$ and $VI$ is given by $\rho=\rho_1>\rho_0$. We have the freedom to choose $f_0, k_0, p_0,\rho_1$ and we have varied them as we moved in the phase space of solutions (i.e. as we changed $T L (\rho)$ to improve numerical convergence. 
At the patch boundaries, we require that the metric $g$ and 2-form $C_{(2)}$ given by \eqref{eqAPP:ansatzxy} and \eqref{eqAPP:ansatzrhoxi} do match, and we also require that their normal derivatives across the patch boundaries also do match. Together with the boundary conditions described above we thus have a well-posed boundary value problem for each of the patches and for the global integration domain.

\section{Energy and expectation value of dual operators}
To find the energy $E$, the holographic stress tensor $T_{ij}$ and the expectation value of holographic dual operators we need the asymptotic Taylor expansion of the fields around $y=1$ up to order $\mathcal{O}\left(1-y\right)^3$. This is given by 
\begin{subequations}\label{asymptotics}
\begin{eqnarray}
f_1{\bigl |}_{y=1}&=& 1+\frac{1}{6} \left[ 2 \alpha_2  Y_2(x)
+\left(12 \alpha_2+2 \beta_1-\beta_1^2+4 \beta_2\right) Y^x_{x\,(2)}(x) \right]  (1-y)^2 +\mathcal{O}\left(1-y\right)^3,
\\
f_2{\bigl |}_{y=1}&=& 1-\frac{1}{36} (2\beta_1-\beta_1^2+4\beta_2) Y_2(x) (1-y)^2  +\mathcal{O}\left(1-y\right)^3 \\
f_3{\bigl |}_{y=1}&=& 1-\frac{1}{18}\left[ (6\alpha_2+2\beta_1-\beta_1^2+4\beta_2) Y_2(x)+
 3 (12\alpha_2+2 \beta_1-\beta_1^2+4 \beta_2) Y^x_{x\,(2)}(x) \right]  (1-y)^2  +\mathcal{O}\left(1-y\right)^3 \\
f_4{\bigl |}_{y=1}&=& 1-\frac{1}{2} \beta_1 \, Y_1(x)   (1-y)
+ \frac{1}{8} \left[ 2 \beta_1 Y_1(x) + \frac{4}{3} (\beta_1+2\beta_2)  Y_2(x) + \beta_1^2\, Y^x_{x\,(2)}(x) \right]  (1-y)^2
+\mathcal{O}\left(1-y\right)^3 \\
f_5{\bigl |}_{y=1}&=& 1-\frac{1}{2} \beta_1\, Y_1(x)   (1-y)  + \frac{1}{4} \left[ \beta_1 Y_1(x) + \frac{2}{3} (\beta_1+2\beta_2)  Y_2(x) + 2\beta_1^2\, Y^{\Omega}_{\Omega \,(2)}(x) \right]  (1-y)^2  
+\mathcal{O}\left(1-y\right)^3 \\
 f_6{\bigl |}_{y=1} &=&
\frac{1}{64}  \left(2-x^2\right) \Bigg[ \sqrt{3} \beta_1 S_x^{(1)}(x) 
-  \left[  \sqrt{3} \beta_1 S_x^{(1)}(x) + \frac{16 \beta_1-17 \beta_1^2+32 \beta_2}{18 \sqrt{2}} S_x^{(2)}(x) \right] (1-y) \Bigg] \nonumber\\
&& +\mathcal{O}\left(1-y\right)^2 \\
f_7{\bigl |}_{y=1}&=& 1+ \frac{1}{3} \left[ \delta_2  Y_2(x) +  \frac{1}{2} \Big(2 \beta_1-\beta_1^2+4 (\beta_2+3 \delta_2 )\Big) Y^x_{x\,(2)}(x)  \right] (1-y)^2
+\mathcal{O}\left(1-y\right)^3 \\
\end{eqnarray}
\end{subequations}
where $Y_\ell(x)$ , with $\ell=0,1,2,\cdots$ are the (regular) scalar harmonics of S$^3$ given by
\begin{equation}
Y_\ell(x)=(-1)^{\ell }\, \frac{\sin[(\ell+1)\theta(x)]}{\sin \theta(x)}\quad\text{with}\quad \theta(x)=4 \arcsin\left(\frac{x}{\sqrt{2}}\right)\,.
\end{equation}
We have chosen the normalisations so that
\begin{equation}
\lim_{x\to1}Y_{\ell}(x)=\ell+1\,,
\end{equation}
which automatically ensures
\begin{equation}
\int_{\Omega_3}Y_{\ell}(\tilde{x})\,Y_{\tilde{\ell}}(\tilde{x})\,\sqrt{\gamma}\,{\rm d}^3\tilde{x}=2\pi^2\delta_{\ell\,\tilde{\ell}}\,,
\end{equation}
with $\gamma_{ab}$ the metric of the unit-radius round S$^3$, $\gamma$ its determinant, and $\tilde{x}$ denoting the coordinates on S$^3$.

In~(\ref{asymptotics}), $S_x^{(\ell)}(x)$ is the first component of the scalar derived vector harmonic $S_a^{(\ell)}$, and $Y^x_{x\,(\ell)}(x)$ and $Y^\Omega_{\Omega\,(\ell)}(x)$ are components of the scalar derived tensor harmonic $Y^a_{b\,(\ell)}(x)$ defined as
\begin{equation}
S_a^{(\ell)}=-\frac{1}{\sqrt{\ell(\ell+2)}}D_a Y_\ell, \qquad\qquad Y^a_{b\,(\ell)}(x)=\frac{1}{\ell (\ell +2)}\,D^a D_b Y_\ell+\frac{1}{3}\,\gamma ^a{}_b Y_\ell\,,
\end{equation}
where $D$ denotes the covariant derivative compatible with the metric $\gamma_{ab}$ of the unit-radius round S$^3$.

In the above expansion we have already imposed the boundary conditions and we stop at $\mathcal{O}(1-y)^2$ that contributes to the energy and expectation value of the dual operator $\mathcal{O}_2$ with lowest conformal dimension (namely $\Delta=2$). In these conditions, the harmonic coefficients depend on four unknown constants $\{ \beta_1, \beta_2, \alpha_2, \delta_2\}$ that are determined only after solving the entire boundary value problem. Physical observables are a function of these constants.   

With the above asymptotic expansion of the fields at the holographic boundary, we can compute the energy and expectation values of dual operators, which depend on (some of) the constants $\{ \beta_1, \beta_2, \alpha_2, \delta_2\}$.  We can do so using the formalism of Kaluza-Klein holographic renormalisation \cite{Kanitscheider:2006zf,Kanitscheider:2007wq} (see also \cite{Skenderis:2006uy,Dias:2015pda,Kim:1985ez,Gunaydin:1984fk,Lee:1998bxa,Lee:1999pj,Arutyunov:1999en,Skenderis:2006di,Skenderis:2007yb}) or the covariant Noether charge formalism (a.k.a the covariant phase method) \cite{wald1993black, iyer1994some, iyer1995comparison, wald2000general, papadimitriou2005thermodynamics, Dias:2019wof}.
The energy of the solution of our black holes (measured with respect to the global AdS$_3\times$S$^3$ solution) is:
\begin{equation} \label{eq:energy}
\frac{EL}{N^2}= -\frac{1}{48} \Bigl[ 24+ 12 \alpha_2+\left(2 \beta_1-\beta_1^2+4 \beta_2\right)\Bigl],
\end{equation}
where, using \eqref{asymptotics}, one gets $\beta_1$ from a nonlinear model fit of the asymptotic value of $f_6$. Then $\alpha_2$ and $\beta_2$ can be found as fitting parameters of the nonlinear model fit of the asymptotic value of $\partial_y^{2} f_1$:
\begin{eqnarray}\label{eq:energyNLmodel}
&& 
f_6 {\bigl |}_{y=1}=-\frac{1}{4}\, \beta_1 \, x \left(2-x^2\right) \left(1-x^2\right),  \nonumber \\
&&
\partial_y^{2} f_1 {\bigl |}_{y=1}=
2 \,\alpha_2+\frac{8}{9} x^2 \left(2-x^2\right) \left(1-x^2\right)^2 \left(2 \beta_1-\beta_1^2+4 \beta_2\right).
\end{eqnarray}

Kaluza-Klein holography also allows us to compute the expectation values of the scalar operators that condensate 
on the boundary theory when the $SO(4)$ R-symmetry of the D1-D5 CFT$_2$ is spontaneously broken down to $SO(3)$. There is an infinite tower of such scalar and R-symmetry current operators but one of them (the scalar operator ${\cal O}_{2}$) has the lowest conformal dimension, $\Delta=2$ (of course, the VEVs of all operators vanish for the $\mathrm{BTZ} \times \mathrm{S}^3\times \mathbb{T}^4$). 
We note that the absence of a $(1-y)\ln(1-y)$ in $f_{4,5,6}$ means that a would be dual scalar operator $\mathcal{O}_1$ with conformal dimension 1 (see \cite{Kanitscheider:2006zf,Kanitscheider:2007wq}) is not excited in our system.
The expectation value (VEV) of the scalar operator ${\cal O}_{2}$ is \cite{Kanitscheider:2006zf}: 
\begin{equation}\label{eq:vevS2} 
\frac{1}{N^2}\left \langle {\cal O}_{2} \right\rangle =\frac{1}{8 \sqrt{6} \,\pi }
\left(\beta_1^2-2 \beta_1-4 \beta_2\right),
\end{equation}
and this is the expectation value that we that we show in Fig.~3 of the main text.  
The reader will note that the energy and $\left \langle {\cal O}_{2} \right\rangle$ don not depend on the parameter $\delta_2$ that appears in \eqref{asymptotics}; $\delta_2$ contributes to the expectation value of a dual R symmetry current operator that has conformal dimension greater than 2 (see \cite{Kanitscheider:2006zf,Kanitscheider:2007wq}). 

To complete the list of thermodynamic observables,  the explicit expressions for the temperature and the entropy are:
\begin{subequations}\label{eq:thermoH}
\begin{align}
        TL&=\frac{3}{2^{8/3} \pi}\,\frac{1}{\rho_0}\;, \\  
        \frac{S}{N^2}&= 8 \rho_0^4 \int_0^1 d\xi \, \xi (1-\xi^2)^2 \, h^2 \, G_1^{1/2} \,G_2 \, {f}_3^{1/2} \, \tilde{f}_4^{1/2} \, {f}_5 {\biggl |}_{\rho=\rho_0}\;.
\end{align}
\end{subequations}

The $SO(3)$ localised black hole solutions must obey the first law of thermodynamics $\mathrm{d}E=T \, \mathrm{d}S$ and we use this fact to monitorize the accuracy of our numerical solutions.
Actually, we can also integrate this first law to compute the energy of the $SO(3)$ localised black holes, instead of using \eqref{eq:energy}. Indeed, recall that computing the energy using \eqref{eq:energy} requires reading the asymptotic value of the second derivative of $f_1$, as described in \eqref{eq:energy}$-$\eqref{eq:energyNLmodel}, and there is an associated error associated to the numerical computation of this second derivative. Alternatively, we can get the energy of the $SO(3)$ localised black holes through the integration of the first law (starting at the AdS$_3$ solution with $E L=-\frac{c}{12}$ and $S=0$ where $SO(3)$ Localised black holes also start existing):
\begin{equation} \label{eq:energy1stLaw}
\frac{E L}{N^2}= \frac{L}{N^2} \int_0 T \mathrm{d}S,
\end{equation}
For most of our $SO(3)$ localised solutions, we find that the energies computed via \eqref{eq:energy} and \eqref{eq:energy1stLaw} agree typically within  $0.01\%$. However, as the first order transition energy, $E_c L/N^2\sim 0.0804$, is approached, this agreement gets less good because the computation \eqref{eq:energyNLmodel} of $\alpha_2$ and $\beta_2$ reading $\partial_y^{2} f_1 {\bigl |}_{y=1}$ becomes considerably harder and less accurate. On the other hand, the energy computed via \eqref{eq:energy1stLaw} only requires the evaluation of the entropy, i.e. of the functions $f_3,\tilde{f}_4,f_5$ at the horizon via \eqref{eq:thermoH}. For this reason, to compute the energy displayed in the plots of the main text we use the integration of the first law \eqref{eq:energy1stLaw}. Further note that we choose not to display $\left \langle {\cal O}_{2} \right\rangle$ for higher energies in Fig.~3 of the main text also because the values of $\alpha_2$ and $\beta_2$ (and thus of $\left \langle {\cal O}_{2} \right\rangle$) become less accurate.

For completeness we also give the energy, entropy, temperature and free energy of the $\mathrm{BTZ} \times \mathrm{S}^3\times \mathbb{T}^4$ black hole (black curves in the plots), of the $U(1)^2$ localised black hole (green curves in the plots) and of the 6-dimensional 
$\mathrm{Schw}_{6} \times \mathbb{T}^4$ black hole (which is a good approximation to the $SO(3)$ localised black holes for small energies; dashed red curves in the plots):
\begin{eqnarray}
&& \hspace{-1cm} \mathrm{BTZ} \times \mathrm{S}^3\times \mathbb{T}^4:  \qquad
\frac{L}{N^2} E_{\hbox{\tiny BTZ}}=\frac{R_+^2}{2}\,,\quad
\frac{1}{N^2}  S_{\hbox{\tiny BTZ}}= 2 \pi  R_+\,,\quad
L T_{\hbox{\tiny BTZ}}= \frac{R_+}{2 \pi }\,,\quad
\frac{L}{N^2} F_{\hbox{\tiny BTZ}}=-\frac{R_+^2}{2}\,; \\
&& \hspace{-1cm} 
U(1)^2: \qquad 
\frac{L}{N^2} E_{\hbox{\tiny $U(1)$}}= 2 \pi ^2 \tilde{T}^2\frac{1-4 \pi ^2 \tilde{T}^2}{\left(1+4 \pi ^2 \tilde{T}^2\right)^2}\,,\quad
\frac{1}{N^2}  S_{\hbox{\tiny $U(1)$}}= \frac{4 \pi ^2 \tilde{T}}{\left(1+4 \pi ^2 \tilde{T}^2 \right)^2} \,,\nonumber \\
&& \hspace{3cm} L T_{\hbox{\tiny $U(1)$}}=  \tilde{T} \,,\quad
\frac{L}{N^2} F_{\hbox{\tiny $U(1)$}}= -\frac{2 \pi ^2 \tilde{T}^2}{1+4 \pi ^2 \tilde{T}^2}\,; \\
&& \hspace{-1cm}
\mathrm{Schw}_{6} \times \mathbb{T}^4:  \qquad
\frac{L}{N^2} E_{\hbox{\tiny Schw$_6$}}= \frac{4 R_+^3}{3 \pi }-\frac{1}{2}\,,\quad
\frac{1}{N^2}  S_{\hbox{\tiny Schw$_6$}}=  \frac{4 R_+^4}{3}\,,\quad
L T_{\hbox{\tiny Schw$_6$}}= \frac{3}{4 \pi  R_+}\,,\quad
\frac{L}{N^2} F_{\hbox{\tiny Schw$_6$}}= \frac{R_+^3}{3 \pi }\,;
\end{eqnarray}
where $R_+=r_+/L$ is the dimensionless horizon radius. 
\section{Phase Diagram in the Canonical Ensemble}

\begin{figure}[th]
\centering
\includegraphics[width=.45\textwidth]{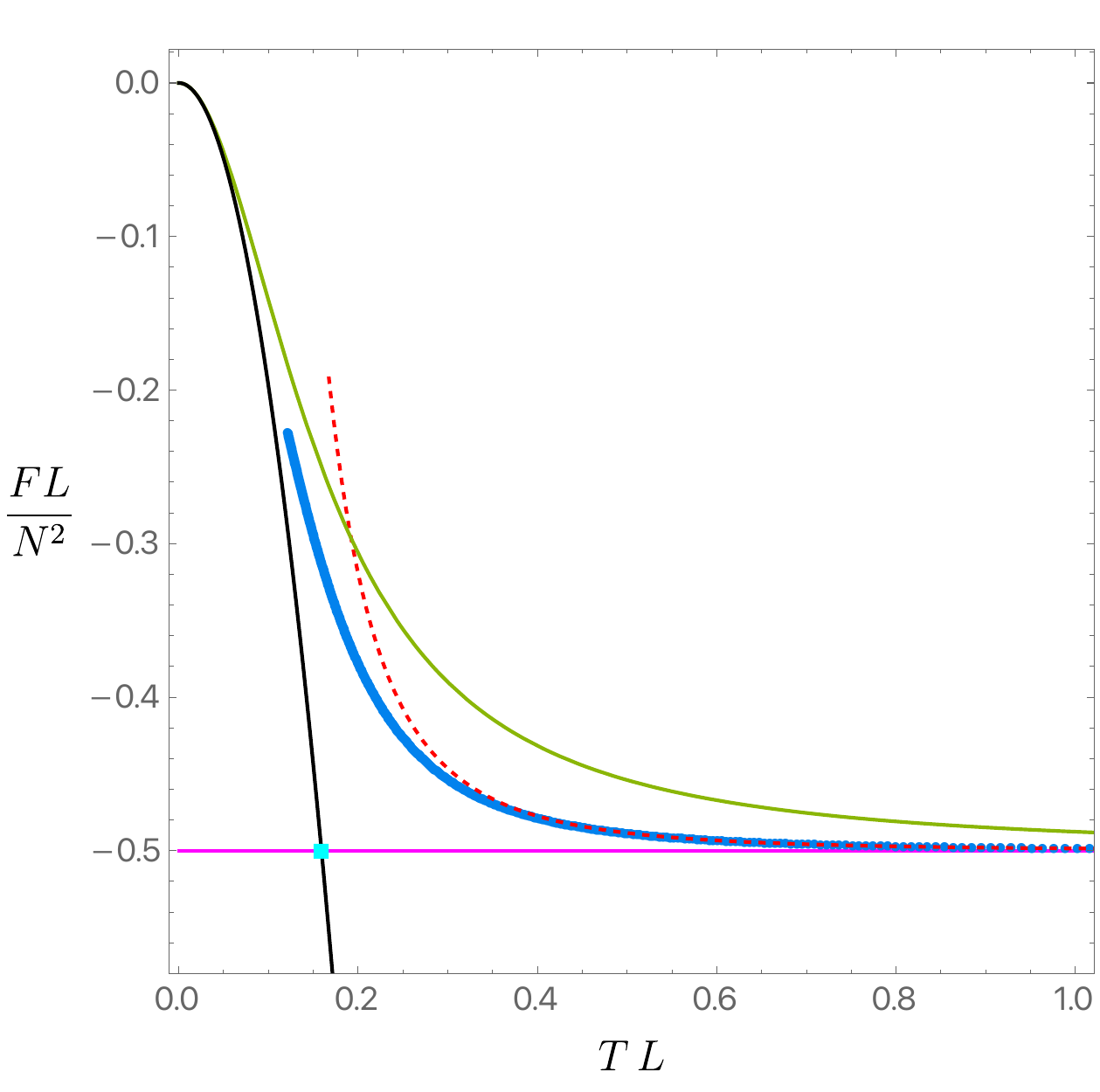}
\caption{Canonical phase diagram: Free energy vs Temperature. Same colour scheme as Fig.~2 of the main text. The magenta line with $F L/N^2=-\frac{1}{2}$ represents thermal $\mathrm{AdS}_3\times \mathrm{S}^3\times \mathbb{T}^4$ and the cyan square at $T_{\mathrm{HP}} L=\frac{1}{2\pi}$ is the Hawking-Page first order transition point. 
}\label{Fig:canonical}
\end{figure}

Below we give a phase diagram of our solutions in the canonical ensemble.  In this ensemble, the temperature is fixed and the solution with the lowest free energy is dominant. Fig.~\ref{Fig:canonical} shows the free energy $FL/N^2$ versus the temperature $TL$. 
Start by noting that, at temperature  $T_{\mathrm{HP}} L=\frac{1}{2\pi}\sim 0.159$ (cyan square), there is a Hawking-Page first-order phase transition between thermal $\mathrm{AdS}_3\times \mathrm{S}^3\times \mathbb{T}^4$ (magenta horizontal line with $F L=-c/12=-\frac{1}{2}N^2$) and $\mathrm{BTZ} \times \mathrm{S}^3\times \mathbb{T}^4$ black holes (black line with $F L=-2\pi^2 T^2 N^2$): thermal AdS is dominant for  $T<T_{\mathrm{HP}}$ but  $\mathrm{BTZ} \times \mathrm{S}^3\times \mathbb{T}^4$ dominates the canonical ensemble for $T>T_{\mathrm{HP}}$. 
Fig.~\ref{Fig:canonical} further shows that our localised solutions with $SO(3)$ symmetry (blue points) have less free energy than the black pole solutions of \cite{Bena:2024gmp} (green curve). However, both these solutions
are, for all temperatures, subdominant with respect to thermal $\mathrm{AdS}_3\times \mathrm{S}^3\times \mathbb{T}^4$  and $\mathrm{BTZ} \times \mathrm{S}^3\times \mathbb{T}^4$.  

\section{No Gregory-Laflamme instability in $\bf{\mathrm{\bf BTZ} \times \mathrm{\bf S}^3\times \mathrm{\bf T}^4}$. Conjectured phase diagram completion.}

An important characteristic of the asymptotically global $\mathrm{AdS}_5\times \mathrm{S}^5$ system of type IIB supergravity is that $\mathrm{Schwarzschild-AdS}_5\times \mathrm{S}^5$ black holes are unstable to the Gregory-Laflamme instability \cite{Gregory:1993vy,Harmark:2002tr,Kol:2002xz,Kudoh:2004hs,Harmark:2007md,Headrick:2009pv,Dias:2015nua} when, roughly, the horizon radius of the black hole is smaller that the radius $L$ of the $\mathrm{S}^5$ (which, as enforced by the equations of motion, is also the $\mathrm{AdS}_5$) \cite{Hubeny:2002xn,Dias:2015pda,Buchel:2015gxa,Dias:2016eto}.
Interestingly, in a phase diagram of solutions, the onset point of this instability then signals a second order phase transition (bifurcation) from the `uniform' black hole to a new family of static black holes that are denoted as 'lumpy' black holes because (unlike $\mathrm{Schwarzschild-AdS}_5\times \mathrm{S}^5$ which have the same horizon topology $\mathrm{S}^4\times  \mathrm{S}^5$ but $SO(6)$ symmetry) these solutions are deformed (i.e. they are non-uniform) along the polar direction of the $\mathrm{S}^5$ \cite{Dias:2015pda}. It turns out that, in the phase diagram, as we keep following this branch of lumpy solutions, the polar deformations increase until the lumpy branch merges (at a conical topology changing point) with the localised black hole branch of solutions with $SO(5)$ symmetry (and broken $SO(6)$ symmetry) and horizon topology $\mathrm{S}^8$ \cite{Dias:2015pda,Dias:2016eto}.  
A similar Gregory-Laflamme instability and associated similar lumpy and $SO(n)$ localised black hole solutions are also present in other asymptotically $\mathrm{AdS}_k\times \mathrm{S}^n$ supergravity systems \cite{Bosch:2017ccw,Dias:2017uyv,Dias:2024vsc}.

Interestingly, the situation is substantially different for the present asymptotically $\mathrm{AdS}_3\times \mathrm{S}^3\times \mathbb{T}^4$ system, although it still has $SO(3)$ localised black holes that we constructed in this manuscript.
Indeed, we find that $\mathrm{BTZ} \times \mathrm{S}^3\times \mathbb{T}^4$ (the `uniform' phase) is {\it not} `Gregory-Laflamme'  unstable for any temperature $T>0$. We can study (analytically) linear mode perturbations (with a Fourier decomposition $e^{-i \omega t}$ along time $t$ which introduces the frequency $\omega$ of the perturbation) of this uniform solution and solve the associated eigenvalue problem for the frequency $\omega$. We further expand all the perturbations in spherical harmonics $Y_{\ell m}$, with $\ell\in\mathbb{Z}^+$ and $|m|\leq\ell$. Details will be given elsewhere, but
 one concludes that the frequencies are quantized as:
 \begin{equation}
 \label{NoGL:freq}
 \omega= -2\pi T\, (2+2p+\ell)\,i\,.
 \end{equation}
The imaginary part of these frequencies is always negative no matter how small the ratio between the horizon radius $r_+$ and the $\mathrm{S}^3$ radius $L$ (recall that $T\propto r_+$ for a BTZ black hole). Therefore, these modes never grow in time and therefore we do not find unstable modes in $\mathrm{BTZ} \times \mathrm{S}^3\times \mathbb{T}^4$.
Consequently, it is now without surprise that we also do not find  $\mathrm{AdS}_3\times \mathrm{S}^3\times \mathbb{T}^4$ `lumpy' black holes that would bifurcate from the $\mathrm{BTZ} \times \mathrm{S}^3\times \mathbb{T}^4$ family and then merge with the localised solutions that we constructed. This is in stark contrast with what happens   in other $\mathrm{AdS}_k\times \mathrm{S}^n$ supergravity systems (e.g. in the $\mathrm{AdS}_5\times \mathrm{S}^5$ system).

\begin{figure}[ht]
\centering
\includegraphics[width=.45\textwidth]{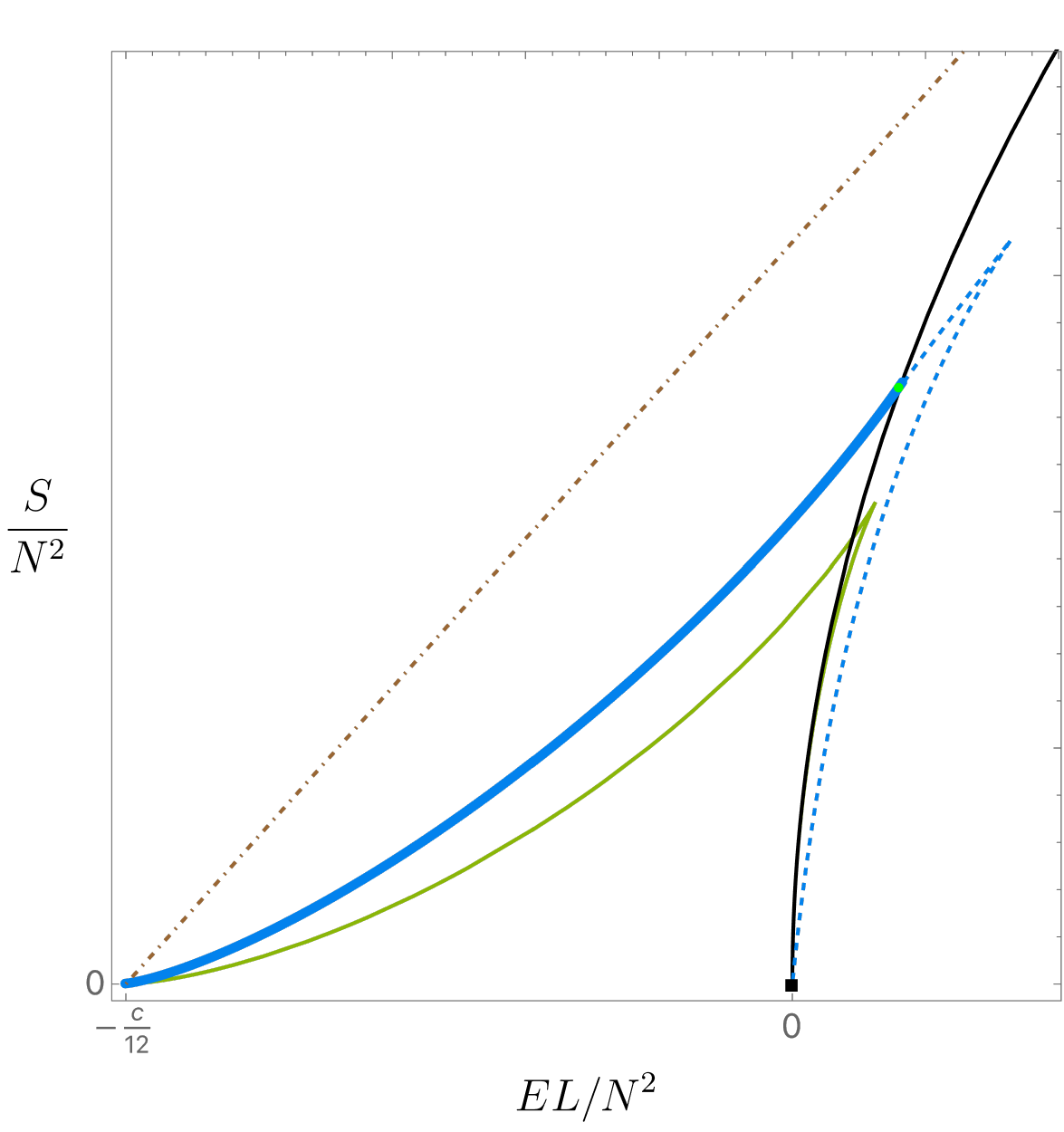}
\hspace{1cm}
\includegraphics[width=.45\textwidth]{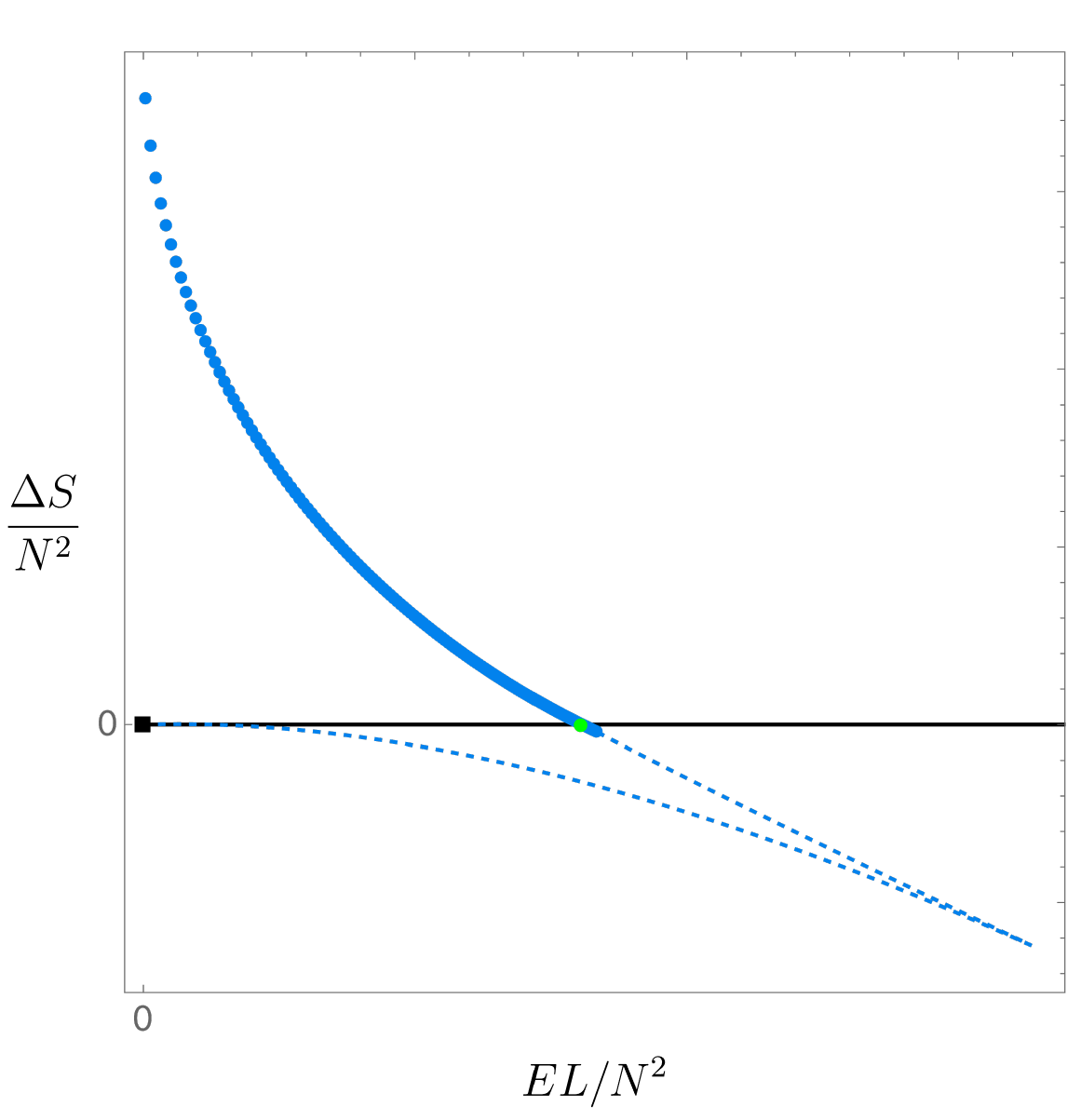}
\caption{Conjectured microcanonical phase diagram: entropy vs energy. The black line is the $\mathrm{BTZ}\times \mathrm{S}^3\times \mathbb{T}^4$ phase, the green curve is the $U(1)^2$ localised phase (the dot-dashed brown line is the `sparseness' bound). The blue dots are the $SO(3)$ localised solutions that we found numerically and the dashed (continous) blue curve with a cusp and ending at the black square with $\{E,S\}=\{0,0\}$ is the conjectured curve that completes the full branch of the $SO(3)$ localised phase. The left panel is the full phase diagram while the right panel focus on the energy range $E\geq 0$ where $SO(3)$ localised solutions can coexist with the BTZ phase and plots the entropy difference $\Delta S$ between the $SO(3)$ (in blue) and BTZ (horizontal black line with $\Delta S=0$) phases.}\label{Fig:APPmicrocanonical}
\end{figure}

Given the absence of the Gregory-Laflamme instability in $\mathrm{BTZ} \times \mathrm{S}^3\times \mathbb{T}^4$ and of would be `lumpy' black holes that branch from the BTZ black hole, one can now conjecture how does the $SO(3)$ localised phase is completed in the microcanonical phase diagram (Fig.~2 of the main text). Recall that we constructed these $SO(3)$ localised black holes only up to an energy $E \sim 0.084 L/N^2$. Fortunately, this was enough to cover the first order phase transition that occurs at  $E =E_c \approx 0.0804 L/N^2$ but it is computationally costly to extend further our numerical construction for higher energies. With the information collected above we can however conjecture the following (see Fig.~\ref{Fig:APPmicrocanonical} which is the conjectured completion of Fig.~2 in the main text). Very much like for the $U(1)^2$ localised solution \cite{Bena:2024gmp} (green curve in Fig.~\ref{Fig:APPmicrocanonical}), where the full solution is analytically available,  one expects $SO(3)$ localised black holes (blue curve) to extend to higher energies until a maximum energy where the cusp is observed in Fig.~\ref{Fig:APPmicrocanonical}. So the `{\it  upper}' $SO(3)$ localised phase should ``end'" at this cusp in Fig.~\ref{Fig:APPmicrocanonical}. But the cusp should also be the starting point of a `{\it lower}' branch of $SO(3)$ localised solutions that should extend all the way down/left to $\{ E,S \}=\{0,0\}$ (black square in Fig.~\ref{Fig:APPmicrocanonical}) where it would meet extremal $\mathrm{BTZ}\times \mathrm{S}^3\times \mathbb{T}^4$ and  the extremal $U(1)^2$ solution of \cite{Bena:2024gmp} with vanishing scalar condensates (this lower $SO(3)$ branch is most probably always below the lower $U(1)^2$ branch as displayed in Fig.~\ref{Fig:APPmicrocanonical}). Note that Fig.~\ref{Fig:APPmicrocanonical} is a {\it sketch} of the conjectured phase diagram completion and thus it is not up to scale. Further note that it does not include multi $SO(3)$ localised black holes (uniform along the torus) neither (single) $SO(3)$ localised black holes that also localize on the $\mathbb{T}^4$ (i.e. that depend non-trivially on the torus coordinates alike those studied in \cite{Dias:2017coo}) that we conjecture to exist in the Discussion section of the main text.  

For completeness, it is natural to conjecture that in the canonical phase diagram of Fig.~\ref{Fig:canonical}, the $SO(3)$ Localised curve (blue disks) should extend all the way up to $F=0$ (at $T=0$) while being always in between the BTZ black curve and the green black pole curve.   
\section{Numerical validation and convergence}

\begin{figure}[ht]
\centering
\includegraphics[width=.45\textwidth]{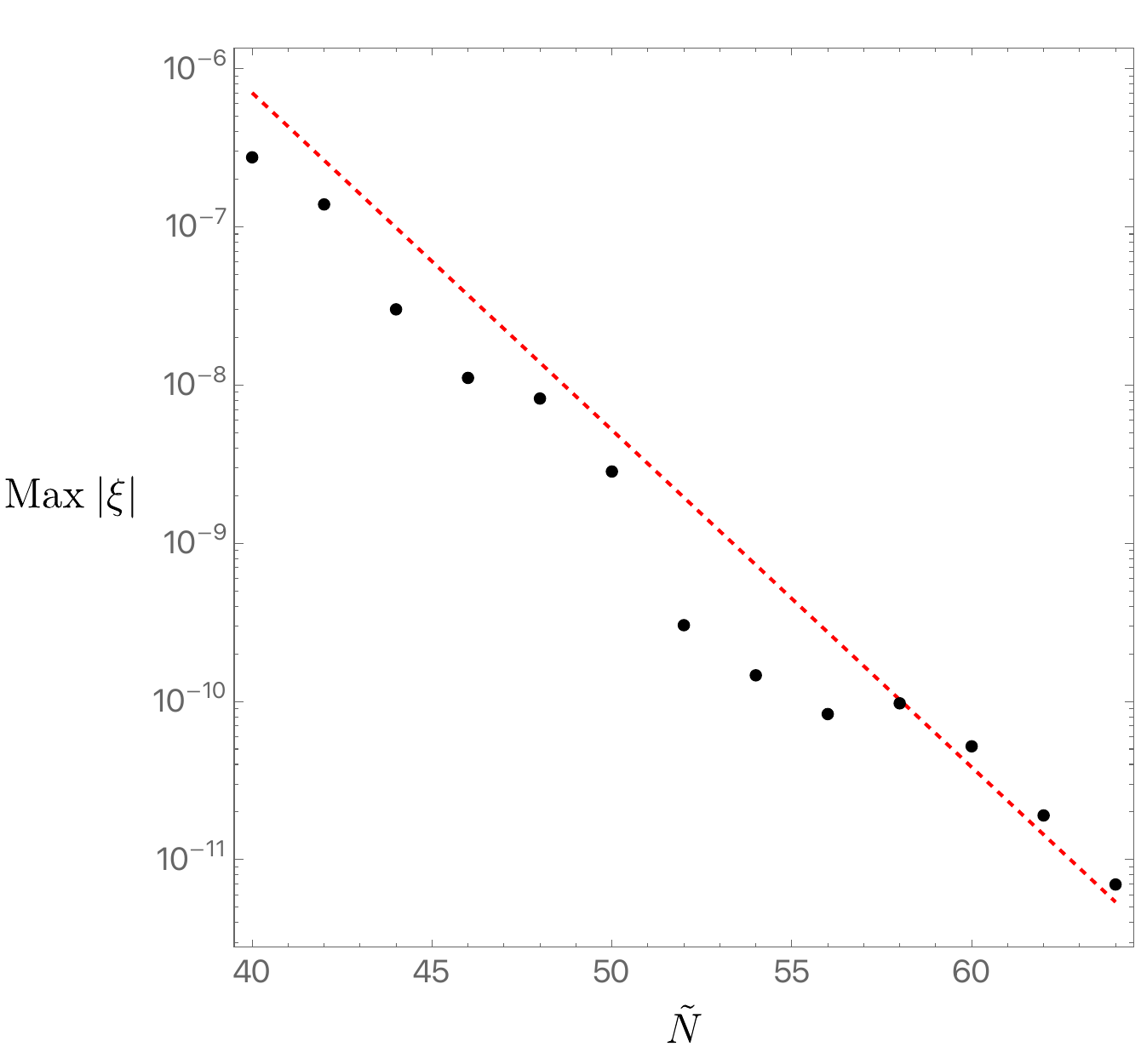}
\caption{Convergence of Max of $|\xi|$  as a function of $\widetilde{N}$ for $\rho_0=1.1$ ($T L= 0.13672$).}
\label{fig:conv}
\end{figure}

In this section we describe some numerical checks that we performed. Within each patch, we have used a $\widetilde{N}\times \widetilde{N}$ size grid. Since our reference metric has been adapted for small black holes (compared with the $S^3$ radius), it is especially difficult to perform accurate numerics on large (energy) black holes. The results we present use up to $\widetilde{N}=60$ at large energies (although we did convergence tests up to $\widetilde{N}=70$). Since we use de Einstein-DeTurck formulation, a quantity that is particularly important to monitor is the norm of the DeTurck vector $\xi$ introduced in equation (4) of the main text. This quantity must vanish for large $\widetilde N$ for any converging numerical discretization method. Since we are using pseudospectral collocation, $|\xi|$ should decrease exponentially in $\widetilde N$ if the solution is sufficiently smooth. In Fig.~\ref{fig:conv} we present convergence test for the maximum (on the numerical grid) of $|\xi|$ and confirm that it shows exponential convergence towards zero.

We can also test our numerics using the fact that the energy extracted at infinity via Kaluza-Klein holography renormalization needs to be equal to the energy extracted by integrating the first law.  In the worst case scenario (large energies), we find that these agree numerically to within $0.1\%$ error. 

\bibliography{refs_ads3xs3}{}
\end{document}